\documentclass[aps,prd,reprint,showpacs,superscriptaddress,
preprintnumbers,nofootinbib]{revtex4-1}
\pdfoutput=1 
\usepackage{amsmath,amssymb,bbm,graphicx,bm,slashed,multirow,color,dcolumn,bbold}
\usepackage[colorlinks=true, pdfstartview=FitV, linkcolor=blue, citecolor=blue, urlcolor=blue]{hyperref}
\usepackage[T1]{fontenc}
\graphicspath{{./Figures/}}
\DeclareMathOperator\ee{e}
\DeclareMathOperator\Tr{Tr}
\DeclareMathOperator\tr{tr}
\DeclareMathOperator\sn{sn}

\DeclareMathOperator\KK{\mathbf{K}}

\DeclareMathOperator\ZZ{\mathbb{Z}}

\newcommand{\M}{\mathcal{M}}
\newcommand{\kkakko}[1]{\left[#1\right]}
\newcommand{\ckakko}[1]{\left\{#1\right\}}
\newcommand{\mkakko}[1]{\left({#1}\right)}
\newcommand{\1}{\mathbb{1}}
\newcommand{\der}{\partial}
\newcommand{\SO}{\text{SO}}
\newcommand{\SU}{\text{SU}}

\renewcommand{\epsilon}{\varepsilon}
\renewcommand{\bar}[1]{\overline{#1}} 
\renewcommand{\Re}{\,\mathrm{Re}}

\def\ba#1\ea{\begin{align}#1\end{align}}
\newcommand{\calO}{\mathcal{O}}
\newcommand{\MM}{M_{0}}
\newcommand{\pp}{\bm{p}}
\newcommand{\dd}{\text{d}}

\begin{document}

\title{Phonons, pions and quasi-long-range order in spatially 
modulated chiral condensates}

\author{Yoshimasa Hidaka} \affiliation{Theoretical Research Division,
Nishina Center, RIKEN, Wako, Saitama 351-0198, Japan}
\author{Kazuhiko Kamikado} \affiliation{Theoretical Research Division, Nishina Center,
RIKEN, Wako, Saitama 351-0198, Japan}
\email[1]{kazuhiko.kamikodo@riken.jp}
\author{Takuya Kanazawa}\affiliation{iTHES Research Group and Quantum Hadron Physics Laboratory,
RIKEN, Wako, Saitama 351-0198, Japan}
\author{Toshifumi Noumi} \affiliation{Theoretical Research Division, Nishina Center, RIKEN, Wako, Saitama 351-0198, Japan}

 \begin{abstract}
   We investigate low-energy fluctuations in the real kink crystal 
  phase of dense quark matter within the Nambu--Jona-Lasinio model. 
  The modulated chiral condensate breaks both the translational symmetry and 
  chiral symmetry spontaneously, which leads to the appearance of phonons 
  and pions that are dominant degrees of freedom in the infrared. 
  Using the Ginzburg--Landau expansion near the Lifshitz point,
  we derive elastic free energies for phonons and pions in dependence on 
  the temperature and chemical potential.   We show  
  that the one-dimensional modulation is destroyed by thermal fluctuations 
  of phonons at nonzero temperature and compute the exponent that 
  characterizes the anisotropic algebraic decay of quasicondensate correlations at long distance. 
  We also estimate finite-volume effects on the stability of the real 
  kink crystal and briefly discuss the possibility of its existence in neutron stars. 
  
 \end{abstract}
 \preprint{RIKEN-QHP-185, RIKEN-STAMP-4}
 \maketitle

 \section{Introduction}
 \label{sec:introduction}

 Revealing the nature of QCD at finite temperature and baryon density has been one of
 the most important challenges in high-energy physics. At low
 temperature and low baryon chemical potential, the ground state of QCD is
 characterized by chiral symmetry breaking and color confinement. 
 Under extreme conditions, QCD is known to exhibit
 novel phenomena, such as deconfinement of color degrees of freedom
 at high temperature and color superconductivity at high baryon
 density. These areas are extensively investigated; 
 see~\cite{Alford:2007xm,Shuryak:2008eq,Fukushima:2010bq} 
 for reviews.
 
 Inhomogeneous phases in QCD have been studied for a long time. Pivotal examples 
 include $p$-wave pion condensation in nuclear matter~\cite{Migdal:1978az}, 
 a chiral density wave at large $N_c$ 
 \cite{Deryagin:1992rw,Shuster:1999tn,Nakano:2004cd,Kojo:2009ha,Kojo:2011cn} and 
 crystalline color superconductivity at high density   
 \cite{Casalbuoni:2003wh,Anglani:2013gfu}, in close analogy to 
 the Fulde--Ferrell--Larkin--Ovchinnikov (FFLO) phases
 of superconductivity~\cite{Fulde:1964zz,larkin1964}.  
 Recently, an exotic phase of dense quark matter 
 in which the chiral condensate forms a spatial soliton lattice has attracted 
 considerable attention \cite{Nickel:2009ke,Nickel:2009wj,
 Carignano:2010ac,Abuki:2011pf,Carignano:2012sx,Carignano:2014jla},
 as reviewed in~\cite{Buballa:2014tba}.  
 Initially, a theoretical breakthrough was made in the study of 
 inhomogeneous chiral condensation in (1+1)-dimensional 
 field theories \cite{Thies:2006ti}. 
 A general complex kink solution to the Ginzburg--Landau (GL) equation for the
 chiral Gross--Neveu model was then discovered 
 \cite{Basar:2008im,Basar:2008ki} and its relevance in the phase diagram 
 of the (chiral) Gross--Neveu model was elucidated \cite{Basar:2009fg}. 
 The issue of inhomogeneous condensation in (3+1)-dimensional theories 
 was revisited by Nickel \cite{Nickel:2009wj} who introduced a novel 
 technique for applying the general solution of the Gross--Neveu model 
 to the Nambu--Jona-Lasinio (NJL) model and thereby pointed out the emergence 
 of a one-dimensionally periodic chiral condensate in the vicinity of the 
 first-order chiral transition line that appears when limiting to a homogeneous 
 condensate. Considering that higher-dimensional modulations are 
 energetically disfavored against one-dimensional modulations 
 \cite{Abuki:2011pf,Carignano:2012sx}, there is a high chance that 
 one-dimensionally modulated chiral condensates may indeed be 
 the ground state of QCD in a certain range of temperature and chemical
 potential. 
 
 So far almost all studies on the inhomogeneous chiral condensation 
 have been limited to the mean-field approximation and inclusion 
 of collective fluctuations of the order parameter is an urgent issue.  
 It has been well known since Landau's seminal work in the 1930s 
 that in three dimensions a one-dimensional 
 spatial order is unstable against thermal fluctuations. 
 This is known as the Landau--Peierls theorem~\cite{landau1969,peierls1934}. 
 The destabilized long-range order leaves its imprint in the power-law decay 
 of the order parameter, characteristic of a phase with quasi-long-range order. 
 This is reminiscent of two-dimensional systems with continuous symmetry, 
 in which a long-range order is prohibited 
 \cite{Mermin:1966fe,Coleman:1973ci} and the Berezinskii--Kosterlitz--Thouless 
 phase emerges at low temperature \cite{berezinskii1971,Kosterlitz:1973xp}.

 In this paper, we elucidate various properties of 
 gapless excitations on a spatially modulated chiral condensate 
 in the NJL model. Our analysis is based on the 6th-order GL expansion 
 of the NJL model near the critical point, for which 
 an analytical solution to the GL equation, called the \emph{real kink crystal} 
 \cite{Basar:2008im,Basar:2008ki,Basar:2009fg}, is available.  
 Because the kink crystal breaks both the translational symmetry along one axis 
 and chiral symmetry, we encounter smectic phonons in addition to ordinary pions. 
 We will derive the elastic free energy of these modes and reveal how they modify 
 the mean-field picture of this phase. 
 While the analysis presented here closely parallels 
 preceding works about fluctuations in liquid crystals 
 \cite{deGennesbook,Chaikin2000,deJeu:2003zz} 
 and in the FFLO phases of fermionic superfluids 
 \cite{radzihovsky2008,samokhin2010,
 samokhin2011,radzihovsky2011a,radzihovsky2011b,PhysRevA.89.013616}, 
 our analysis has a specific focus on quantitative understanding of 
 those gapless modes in the context of dense QCD. 

 The outline of this paper is as follows. 
 In Sec.~\ref{sec:general-discussion}, we develop a general argument 
 for low-energy fluctuations over a one-dimensionally modulated order parameter. 
 In Sec.~\ref{sec:real-kink-crystal}, the GL-expanded NJL model is analyzed. 
 We first derive the free energy for the translational phonon mode  
 and argue that thermal fluctuations of phonons wash out the modulated 
 condensate and lead to a phase with quasi-long-range order. Next 
 we derive the free energy of pions. 
 Finally, Sec.~\ref{sec:conclusion} is devoted to concluding 
 remarks. Some technicalities are summarized in the Appendixes.

 \section{Symmetry consideration}
 \label{sec:general-discussion} 
  
  When a chiral condensate is modulated along one dimension, 
  the phonon mode ($u$) appears as the Nambu--Goldstone (NG) mode of 
  translational symmetry breaking,
  in addition to pions ($\pi$).
  In the real kink crystal phase discussed later, 
  the vectorial isospin symmetry $\SU(2)_{V}$ 
  is unbroken and there is no mixing between phonons and pions.  
  Therefore, the symmetry breaking pattern reads\footnote{The free energy has no Lorentz symmetry because it is broken by the existence of matter. Therefore, we do not take into account the Lorentz symmetry breaking as the breaking pattern.}
    \begin{equation}
  \begin{split}
   \bm{\mathrm{R}}^{3}\rtimes \SO(3)\to      
   \big[ \bm{\mathrm{R}}^{2}\rtimes \SO(2) \big] 
   \times \big[\text{discrete symmetry}\bigr]
  \end{split}
  \label{eq:symmm}
  \end{equation}
  in addition to the chiral symmetry breaking pattern, $\SU(2)_R\times \SU(2)_L\to \SU(2)_V$. 
  Here ${\bm{\mathrm{R}}}^{d}$ and $\SO(d)$ denote the $d$-dimensional translational and rotational symmetry groups, respectively. 
  The discrete symmetry includes a discrete translational symmetry along the modulated direction of the condensate as a subgroup, which is a remnant of 
  the translation ${\bm{\mathrm{R}}}$ in this direction.  Other elements of the discrete group depend on the shape of the kink crystal\footnote{In general, real kink crystals can be
  classified by the Frieze group.  The real kink crystal discussed in this paper has the $\ZZ \rtimes \ZZ_2$ symmetry, where $\ZZ$ and $\ZZ_2$ represent 
 the glide reflection symmetry and  the reflection symmetry at a certain vertical line, respectively.
  }. 

  In this symmetry breaking pattern, two rotational and one translational symmetries are spontaneously broken.
We note that there appears no gapless mode associated with the broken rotational symmetry. In general, the number of NG modes does not coincide with that of broken global spacetime symmetries \cite{Low:2001bw,Watanabe:2013iia,Hayata:2013vfa}.
  
  In the remainder of this section, we present a general discussion on the effective
  low-energy theory of phonons (see also~\cite{landau1969,deGennesbook,Chaikin2000,Leutwyler:1996er,Hidaka:2014fra}).
  Let us consider a theory with
  the free energy $F[\phi]=\int \dd^3x~\mathcal{F}(\phi,\der \phi)$ and assume that
  $\langle\phi\rangle=\phi_0(\bm{x})$ is a static solution minimizing $F[\phi]$.  In the
  following, we assume that the free energy density $\mathcal{F}$ respects rotational and translational
  symmetries. 
  If $\phi_0$ is modulated in one direction and is homogeneous in the
  transverse directions, one can write it as \mbox{$\phi_0(\bm{x})\propto f(\bm{q}\cdot\bm{
  x})$}, where $\bm{q}$ is a vector parallel to the modulated direction and
  $f(\cdot)$ is a dimensionless function.  Now we consider a translational fluctuation corresponding to the phonon around this solution, \mbox{$\phi(\bm{x})\propto
  f(\bm{q}\cdot\bm{x} + q u(\bm{x}))$} with $q=|\bm{q}|$. Plugging this function into $F[\phi]$ and
  expanding in powers of $u$ and $\nabla u$, one obtains the effective
  theory for the $u$ field.
  
  Rotational symmetry of the original free energy implies that $f(\bm{q}\cdot\bm{
  x})$ and $f(\bm{q'}\cdot\bm{x})$, with a rotated vector $\bm{q'}$, have
  the same value of $F[\phi]$. Writing $\bm{q'}\cdot\bm{x}= \bm{q}\cdot\bm{
  x}+(\bm{q'}-\bm{q})\cdot\bm{x}$, it follows that the fluctuation
  $qu(\bm{x})=(\bm{q'}-\bm{q})\cdot\bm{x}$ does not cost any additional
  free energy. This imposes a constraint on the form of the effective theory.  
  Suppose $\bm{q}$ points in the $z$ direction, with no loss of generality. 
  Under a rotation by $\theta$ about the $y$ axis, the
  gradients of $u$ are given by
  \ba
    \der_z u = - 1 +\cos\theta\,, \quad \der_x u = \sin \theta\,,
  \ea
  hence $(1+ \der_z u)^2+(\nabla_\perp u)^2$ is invariant under the
  rotation. Then the effective theory of $u$ invariant under rotation can be
  constructed as a function of $(1+ \der_z u)^2+(\nabla_\perp u)^2
  = 2\der_z u + (\nabla u)^2+1$ and higher-order derivative terms such 
  as $\nabla_\bot^2 u$. 
  
  A total derivative term $\der_z u$ would be allowed in the effective theory 
  if we only require the stationary condition for $F$ under 
  a variation of $\phi$ with fixed boundary conditions. However, 
  global minimization of $F$ excludes this term, which can be seen as follows.  
  Consider a fluctuation $u=\epsilon z$ over the ground state $\phi_0(z)$. 
  It corresponds to a dilatation $z\to (1+ \epsilon)z$.  
  In the presence of a term $\propto A\der_z u$ in $\mathcal{F}$, 
  this $u$ will generate an energy shift 
  \ba
    \label{kink_EFT_enegy}
    \Delta\mathcal{F}
    &= \epsilon A+\mathcal{O}(\epsilon^2)
    \,,
  \ea
  indicating that there exists a lower-energy direction when $A\neq 0$.
  In other words, we have $A=0$
   when the condensate is the ground state. 
  
  This observation imposes a constraint that $\mathcal{F}$ should depend
  on $2\der_z u + (\nabla u)^2$ at least quadratically. Thus we conclude
  that the free energy density for the phonon assumes a form \ba
  \mathcal{F} = B \kkakko{\der_z u + \frac{1}{2} (\nabla u)^2 }^2 + C
  (\nabla_\bot^2 u)^2\,, \label{eq:uuu} \ea at the leading order of the
  derivative expansion, with low-energy constants $B$ and $C$. In
  deriving \eqref{eq:uuu} we exploited the fact that the term
  $(\partial_z u)(\nabla_\bot^2u)$ is prohibited if we assume parity
  invariance at low energy (see Appendix \ref{app:nok3term} for more
  details).  Here we emphasize that the absence of $(\nabla_\bot u)^2$
  in $\mathcal{F}$ makes the dispersion of $u$ strongly anisotropic.
  
  Such an anisotropic dispersion does not appear when the rotational
  symmetry is \emph{explicitly} broken, just as in QCD under external
  magnetic fields.  This is shown in Appendix \ref{app:generaleft}.

 \section{Real kink crystal}
 \label{sec:real-kink-crystal}
 
 In this section, we study the long-wavelength fluctuations of the real
 kink crystal chiral condensate within the NJL model in the chiral
 limit.  The real kink crystal is known to be energetically favored over
 a spatially homogeneous chiral condensate in a certain range of
 temperature and chemical potential \cite{Nickel:2009wj}.  We analyze
 the GL free energy of the NJL model near the Lifshitz point and derive
 the elastic free energy of phonons and pions on the kink crystal on the
 basis of Bloch's theorem for particles moving in a periodic potential.
 For simplicity we will not consider dynamical aspects of those
 fluctuations and also ignore gapped fluctuations.

  \subsection{Phase diagram}
  \label{sec:phase-diagram}
  
  According to Nickel's work \cite{Nickel:2009ke}, the GL expansion for 
  the $(3+1)$-dimensional NJL model in the chiral limit, 
  to sixth order, reads 
   \ba
    \begin{split}
    \Omega_{\rm GL}[M(\bm{x})] & = \alpha_2 M^2 
    + \alpha_4 \ckakko{M^4+(\nabla M)^2} 
    \\    
    & \quad + \alpha_6 \ckakko{ 2M^6 + 10M^2(\nabla M)^2+(\Delta M)^2 } \,. 
    \end{split}
    \label{eq:GL_NJL4}
   \ea
   Here the real-valued field $M(\bm{x})$ represents the chiral condensate. For simplicity, 
   the pionic condensates $\langle\bar\psi i\gamma_5\tau^a\psi\rangle$ are suppressed 
   at this stage. The neutral pion fluctuation will be later incorporated in Sec.~\ref{sec:pions-kink-crystal}, 
   and a fully isospin-symmetric form of the GL expansion is presented in Appendix \ref{sec:gl-expansion-with}.   
   
   The coefficients in the GL expansion \eqref{eq:GL_NJL4} are given as functions of
   temperature ($T$) and chemical potential ($\mu$) \cite{Nickel:2009ke} 
   \ba 
     \begin{split}
     \alpha_2 &= \frac{\alpha'_2}{2} + \frac{1}{4G} \,,\; \alpha_4 =
     \frac{\alpha'_4}{4}\,,\;  \alpha_6 = \frac{\alpha'_6}{12} \,,
     \\
     \alpha'_n & \equiv  (-1)^{\frac{n}{2}} 4 N_c N_f T \sum_n 
     \int_{\rm reg} \frac{\dd^3p}{(2
     \pi)^3}
     \frac{1}{\left[(\omega_n+i\mu)^2 + p^2 \right]^{n/2}}  \,,
     \end{split}
     \label{eq:alphas}
   \ea
   where $G$ is the four-fermion coupling in the NJL model and $N_c$ and
   $N_f$ are the numbers of colors and flavors, respectively. The momentum 
   integrals for $\alpha_2'$ and $\alpha_4'$ are ultraviolet divergent 
   and must be regularized with some regularization scheme. 
   This GL expansion is valid near the QCD critical point at which $\alpha_2 = \alpha_4=0$. 
   We require $\alpha_6>0$ for stability. 

   To find the correct ground state of the GL free energy, we
   have to solve the GL equation:
   \ba
     \frac{\delta}{\delta M(\bm{x})} \int \dd^3y~ 
     \Omega_{\rm GL}[M(\bm{y})] =0\;.
     \label{eq:GL_equation}
   \ea
   If we assume a one-dimensional modulation in the $z$ direction, the GL
   equation reduces to that of the GN$_2$ model, which has a family of 
   solutions \cite{Basar:2008im,Basar:2008ki,Basar:2009fg} given by
   \ba
     \MM(z) & = q \sqrt{\nu} \sn(q z; \nu) \,,
     \label{eq:Mzero}
   \ea
   where $q$ is a function of $\nu$ through the relation
   \ba
     q^4 + \frac{\nu+1}{\nu^2+4\nu+1}\frac{\alpha_4}{\alpha_6}q^2 
     + \frac{1}{\nu^2+4\nu+1}\frac{\alpha_2}{\alpha_6}=0\,. 
     \label{eq:cons1}
   \ea
   One can easily check that $M_0$ with \eqref{eq:cons1} satisfies the
   GL equation \eqref{eq:GL_equation}.%
   \footnote{The most general solution to \eqref{eq:GL_equation} 
   is given by a quasi-periodic function written by the Riemann 
   theta function with genus $g=3$ \cite{Takahashi2013x}, which is 
   considerably more complicated than the elliptic function, \eqref{eq:Mzero}. 
   In the following, we assume that \eqref{eq:Mzero} is energetically 
   favored over higher genus functions, and leave a more comprehensive 
   analysis to future work. 
   We thank D.~A.~Takahashi for useful comments on this point.} 
   The elliptic parameter $\nu\in [0,1]$ 
   is not determined by the GL equation alone 
   and must be fixed from the requirement of lowest energy per period. 
   $\nu$ controls the shape of the 
   solution; $\sn(z;0)=\sin z$ and $\sn(z;1)=\tanh z$. We can identify
   the solution $\sn(z;1)$ as the homogeneous solution because
   thermodynamically $\tanh z$ is equivalent to a spatially homogeneous
   configuration. 
   Let $L$ be the period of $M_0$ 
   and $Q$ the wave number of $M_0$, respectively given by
   \ba
     L \equiv \frac{4 \KK(\nu)}{q} \quad \text{and} \quad Q \equiv \frac{2\pi}{L}\,,
     \label{eq:defLQ}
   \ea
   where \mbox{$\KK(\nu)\equiv\int_0^{\pi/2}\dd t \, (1-\nu \sin^2 t)^{-1/2}$} 
   is the complete elliptic integral of the first kind. 
   In what follows, we denote the average of a periodic function as
   \ba
       \oint F \equiv \frac{1}{L}\int_{0}^{L}\dd z F(z) \,.
   \ea
   
   To map out the phase diagram near the tricritical point,  we need to evaluate 
   \eqref{eq:alphas} for given $T$ and $\mu$.  In this work, 
   the Matsubara sum in \eqref{eq:alphas} was done analytically, whereas
   the momentum integral was regularized by a three-momentum cutoff $\Lambda$. 
   (See Appendix \ref{app:propertime} for additional comments on the choice of 
   regularization.) In this work, we have used the parameter set 
   in \cite{Hatsuda:1994pi}: $\Lambda=632$ MeV and $G \Lambda^2=2.173$. 
   The value of $\nu$ at a given $(T,\mu)$ was numerically determined  
   through minimization of the GL free energy per period 
   \ba 
     \oint  \Omega_{\rm GL} [M_0(z)]  \,. 
     \label{eq:free_energy}
   \ea 
   Then $q$ immediately follows from $\nu$ via \eqref{eq:cons1}. 
  
\begin{figure}[t]
     \centering
     \includegraphics[width=.49\textwidth]{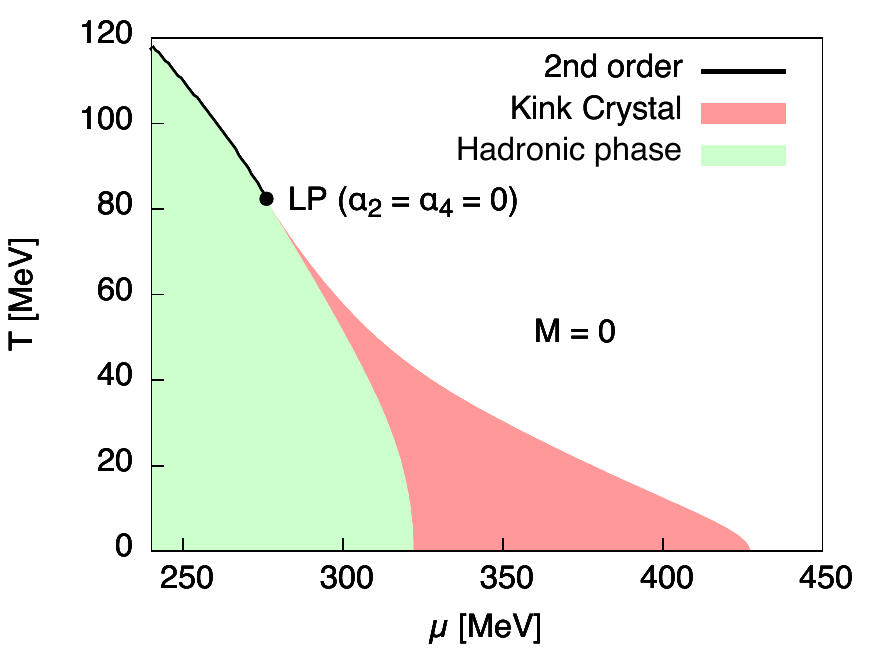}
     \caption{\label{fig:phase_withinhomo}
       Phase diagram of the NJL model in the chiral limit with 6th order GL
       expansion. The Lifshitz point is located at $(T,\mu)= (81.4,276.4)$ [MeV]. 
     }
  \end{figure}

   In Fig.~\ref{fig:phase_withinhomo}, we show the resulting 
   phase diagram of the GL-expanded NJL model \eqref{eq:GL_NJL4}. 
   Strictly speaking, the GL analysis is only valid in a region with a small 
   order parameter and small spatial variation, so 
   the phase structure shown here is only intended to be a qualitative guide.  
   In the red region, the kink crystal chiral condensate $(0<\nu<1)$ is
   energetically favored, while in the green region, the homogeneous 
   chiral condensate ($\nu =1$) develops. The three phases (real
   kink crystal, homogeneous and symmetric phases) meet at the Lifshitz
   point, $(T,\mu)= (81.4,276.4)$, at which $\alpha_2 = \alpha_4=0$. The
   real kink crystal phase is very narrow near the Lifshitz point and is
   almost invisible in the plot. The phase structure from our numerical
   calculation is consistent with analyses with a 
   non-expanded effective potential \cite{Nickel:2009wj} except that the
   real kink crystal phase in Fig.~\ref{fig:phase_withinhomo} 
   appears to be broader at low temperature. The phase boundary 
   in our results are consistent with Nickel's observation \cite{Nickel:2009ke} 
   that the real kink crystal phase is favored if $\alpha_2>0$ and 
   \ba
     - \sqrt{\frac{54}{5}\alpha_2\alpha_6} < \alpha_4 < -2 \sqrt{\alpha_2 \alpha_6} \,. 
   \ea
   
   \begin{figure}[t]
     \centering 
     \includegraphics[width=.49\textwidth]{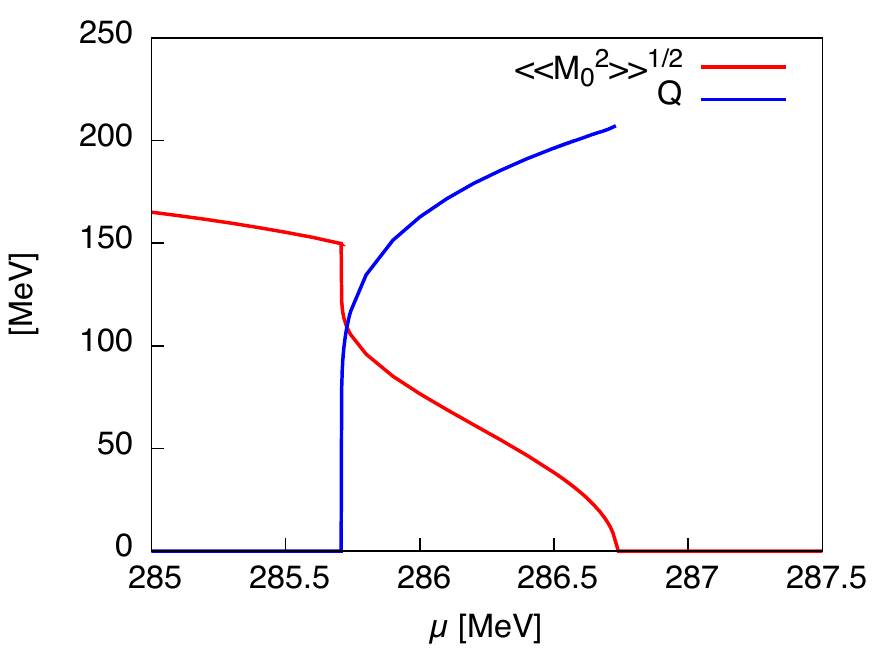}
     \includegraphics[width=.49\textwidth]{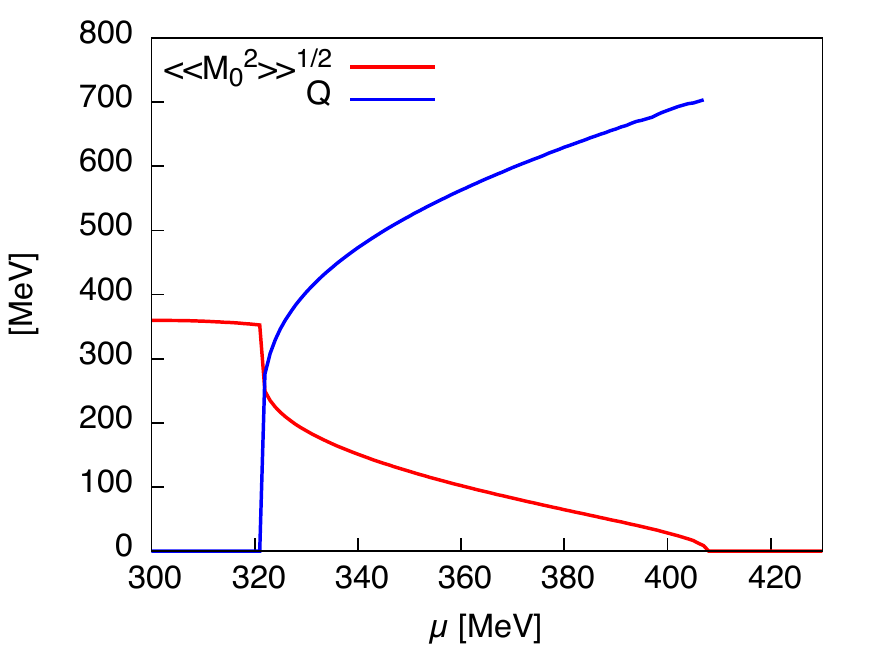}
     \caption{\label{fig:order_parameters}
       The average magnitude of $M$ and its wave number $Q$, 
       at $T = 70$ MeV ({\bf top}) and $T = 10$ MeV ({\bf bottom}). 
     }
   \end{figure}
   In Fig.~\ref{fig:order_parameters}, we plot the root-mean-square 
   condensate 
   \ba
     \langle\!\langle M_0^2 \rangle\!\rangle^{1/2} \equiv \sqrt{\oint M_0(z)^2}  
   \ea
   and the wave number $Q$ of the real kink crystal at $T=70$ MeV
   and $T=10$ MeV. It is observed that 
   both phase transitions (from the kink crystal phase
   to the homogeneous and to the symmetric phase) are of second order:  
   At the transition with lower $\mu$, the wave number gradually 
   rises from zero, implying the formation of widely separated domain walls. 
   On the other hand, at the transition with higher $\mu$, the amplitude of $M$ 
   vanishes smoothly with keeping a nonzero wave number. This behavior 
   is consistent with the preceding work with a non-expanded potential
   \cite{Nickel:2009wj}.

   \subsection{Phonons}
   \label{sec:phonon-mode}

   We would like to derive the elastic free energy of phonons originating 
   from the spontaneous breaking of 
   translation symmetry in the real kink crystal phase.  
   Let us substitute 
   \ba
     M(\bm{x}) & = M_0 \big(z+u(\bm{x})\big)
     \notag
     \\
     & = M_0(z) + M_0'(z) u(\bm{x}) + \frac{1}{2}M_0''(z)u(\bm{x})^2 + \cdots 
     \notag
   \ea
   into \eqref{eq:GL_NJL4} and expand in $u$, dropping 
   total derivatives. Then 
   \ba
    \begin{split}
    & \Omega_{\rm GL}[M(\bm{x})] 
    \\
    &\quad = \Omega_{\rm GL}[M_0(z)] +  \frac{f_1(z)}{2}(\der_z  u)^2
    +\frac{f_2(z)}{2}(\der_z^2  u)^2 
    \\
    & \qquad + \frac{g_1(z)}{2}
    (\nabla_\perp u)^2+ \frac{g_2(z)}{2}(\nabla_{\perp}^2u)^2 
    \\
    & \qquad + h_1(z) (\der_z u)(\nabla_{\perp}^2 u) 
    \\
    & \qquad 
    + {h_2}(z) (\der_z^2 u) (\nabla_{\perp}^2 u)  
    + \calO(u^3)\,, 
    \end{split}
    \label{eq:OGLeq}
   \ea
   where $\nabla_\perp\equiv(\der_x,\der_y)$ is a transverse derivative and  
   $f_1$, $f_2$, $g_1$, $g_2$, $h_1$, $h_2$ are defined as
   \ba
     \begin{split}
     f_1(z) &= 2 (\alpha_4+10\alpha_6M_0^2) (M_0')^2 
     \\
     & \qquad + 4\alpha_6 \ckakko{(M_0'')^2-2M_0'M_0'''} \,,
     \\
     f_2(z) &= 2 \alpha_6 (M_0')^2 \,,
     \\
     g_1(z) &= 2 (\alpha_4+10\alpha_6M_0^2) (M_0')^2 - 4
     \alpha_6 M_0'M_0''' \,,
     \\
     g_2(z) &=2\alpha_6 (M_0')^2 \,, 
     \\
     h_1(z) & = 4 \alpha_6    M_0'M_0''
     \;,\quad  h_2(z) = 2 \alpha_6 (M_0')^2 \;.
     \end{split}
   \ea
   
   \begin{figure}[t]
     \centering 
     \includegraphics[width=.49\textwidth]{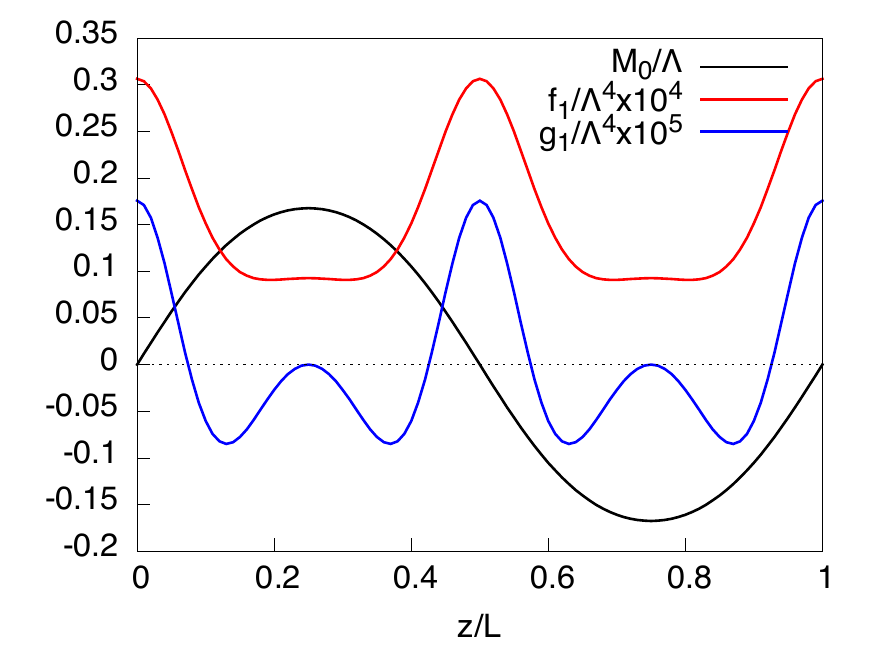}
     \includegraphics[width=.49\textwidth]{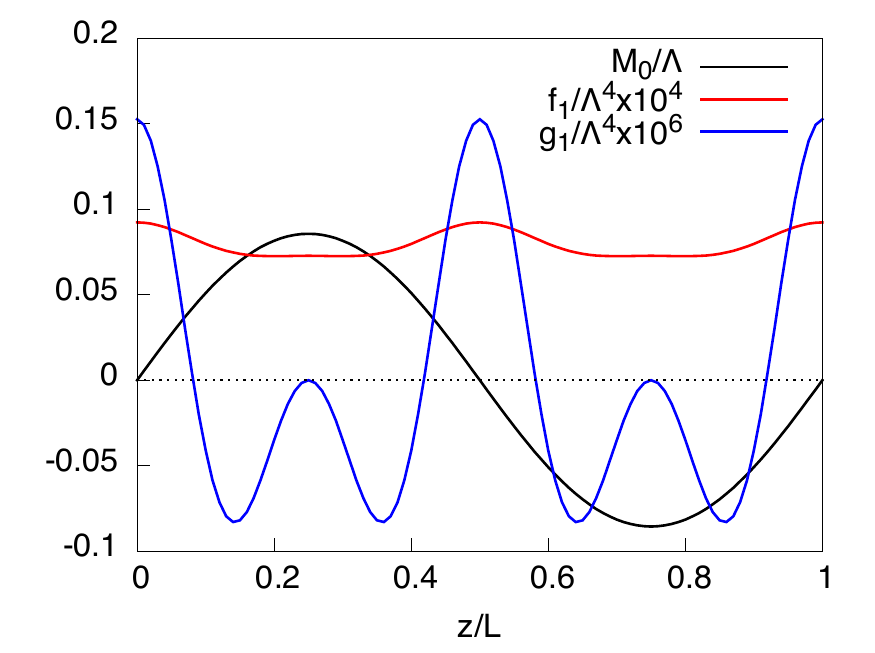}
     \caption{\label{fig:funcs} 
     The order parameter $M_0$ and the coefficient functions $f_1$
     and $g_1$ over one period 
     at $(T,\mu) = (70 , 286.0)$ [MeV] ({\bf top}) and $(T,\mu)
     = (70 , 286.5)$ [MeV] ({\bf bottom}). All functions are measured 
     in units of $\Lambda$. 
     The period $L$ is $7.71$ fm ({\bf top}) and $6.40$ fm ({\bf bottom}), 
     respectively. 
     }
   \end{figure}
   
   Figure \ref{fig:funcs} displays 
   the order parameter and some of the coefficient functions at 
   $(T,\mu) = (70 , 286.0)$ [MeV] and $(T,\mu) = (70 , 286.5)$ [MeV]. 
   Obviously the functions share the same period with the order parameter.  
   Note that $(\nabla_\bot u)^2$ is present in 
   \eqref{eq:OGLeq} even though the general theory in Sec.~\ref{sec:general-discussion} 
   suggests that this term should be absent in the effective action of low-energy modes. 
   This is not a contradiction: since the expansion \eqref{eq:OGLeq} includes 
   functions $f_1$, $g_1$, etc., that vary over the microscopic length scale $L$, we cannot 
   immediately infer the dispersion of modes with wavelength much longer 
   than $L$ from there.
   
   As emphasized in Sec.~\ref{sec:general-discussion} the free energy of a modulated 
   condensate must be invariant under a spatial rotation. This can be checked 
   for \eqref{eq:OGLeq} as follows. An infinitesimal 
   rotation about $y$ axis is equivalent to a displacement field 
   $u(\bm{x})= - \sin\theta\,x+(1-\cos\theta)z$ with $|\theta|\ll 1$. 
   Plugging this into \eqref{eq:OGLeq} and expanding in $\theta$, we find 
   the leading term to be $\frac{1}{2} g_1(z) \theta^2$. Then,  
   for the rotational symmetry to be preserved, the average of $g_1$ must vanish: 
   \ba
     \oint g_1(z) = 0\;.
     \label{eq:ava_g_1}
   \ea 
   A direct proof of this equality based on the minimization of energy 
   is given in Appendix~\ref{sec:proof_g1=0} for completeness.  
   We remark that \eqref{eq:ava_g_1} is not automatically ensured 
   by the GL equation \eqref{eq:GL_equation} 
   [or equivalently \eqref{eq:cons1}] alone---we must minimize 
   the energy per period, to have \eqref{eq:ava_g_1} satisfied. 
   It will be shown below that the property \eqref{eq:ava_g_1} is 
   instrumental in making the dispersion of phonons anisotropic, 
   in accordance with the general argument in Sec.~\ref{sec:general-discussion}.    
       
   To evaluate the low-energy 
   phonon fluctuation on the real kink crystal, 
   let us consider the eigenvalue equation
   \ba
     E  u = \frac{\delta{F[u]}}{\delta u} 
   \ea
   with
   \ba
     \begin{split}
     F[u] & \equiv \int \dd^3 x \;\Bigl[ \frac{f_1(z)}{2}(\der_z  u)^2 
     +\frac{f_2(z)}{2}(\der_z^2  u)^2 
     \\
     & \quad + \frac{g_1(z)}{2}
     (\nabla_\perp u)^2+ \frac{g_2(z)}{2}(\nabla_{\perp}^2u)^2 
     \\
     & \quad 
     +  h_1(z) (\der_z u)(\nabla_{\perp}^2 u) 
     + h_2(z) (\der_z^2 u)(\nabla_{\perp}^2 u) \Bigr]\;.
     \end{split}
     \label{eq:Hamiltonian}
   \ea
   The eigenvalue equation in the explicit form reads
   \ba
     \begin{split}
    H_u u = E u 
     \end{split}
     \label{eq:schrodinger_eq}
   \ea
   with
   \ba
    \begin{split}
    H_u&\equiv- \der_z (f_1 \der_z)+\der_z^2 (f_2
     \der_z^2) - g_1 \nabla_{\perp}^2 + g_2 \nabla_{\perp}^4   
     \\
     &\quad- (\der_z h_1)  \nabla_{\perp}^2 + \nabla_{\perp}^2 
     \ckakko{h_2,\der_z^2}_+ \,,
     \label{eq:Hu}
   \end{split}
   \ea
   where $\ckakko{}_+$ denotes the anti-commutation relation and
   $\ckakko{h_2,\der_z^2}_+ u = h_2 \partial_z^2 u + \partial_z^2(h_2
   u)$. As the operator $H_u$ acting on $u$ is real and Hermitian, the
   eigenvalue $E$ is real, and we can elevate $u$ to a complex-valued function 
   without changing the eigenvalues. 
   We note that $E$ itself does not give the dispersion relation of the phonon, but is rather related to  the phonon susceptibility.
   The dispersion relation can in principle be obtained from the time evolution equation, but it is complicated because of the mixing with hydrodynamic modes (see, e.g., \cite{Chaikin2000}) and 
   will not be discussed further in this paper.

   Since all the coefficient functions are periodic functions sharing
   the same period, it follows from Bloch's theorem that 
   $u$ can be decomposed into a plane wave and a periodic function,
   \ba
     u(\bm{x}) = \ee^{ i\bm{k}_{\perp}
     \cdot \bm{x}_{\perp}}
     \ee^{ ik_{z} z} \phi(z) \,,
     \label{eq:Bloch}
   \ea
   where $k_{\perp}$ is the momentum in transverse directions,  
   $k_{z}$ is the so-called \emph{crystal momentum}, and
   $\phi(z)$ is a periodic function, viz.~$\phi(z+L)=\phi(z)$. Substituting
   \eqref{eq:Bloch} into \eqref{eq:schrodinger_eq} yields an 
   eigenvalue equation for $\phi$, which we have solved numerically 
   by way of a Fourier decomposition $\phi(z) = \sum_{n=-n_{\max}}^{n_{\max}}
   \phi_n\ee^{i n Qz }/\sqrt{L}$ with $n_{\max}=20$.  
   To see convergence, we have increased $n_{\max}$ up to $30$ 
   and confirmed that the results are unchanged.

   \begin{figure}[t]
     \centering
     \includegraphics[width=.49\textwidth]{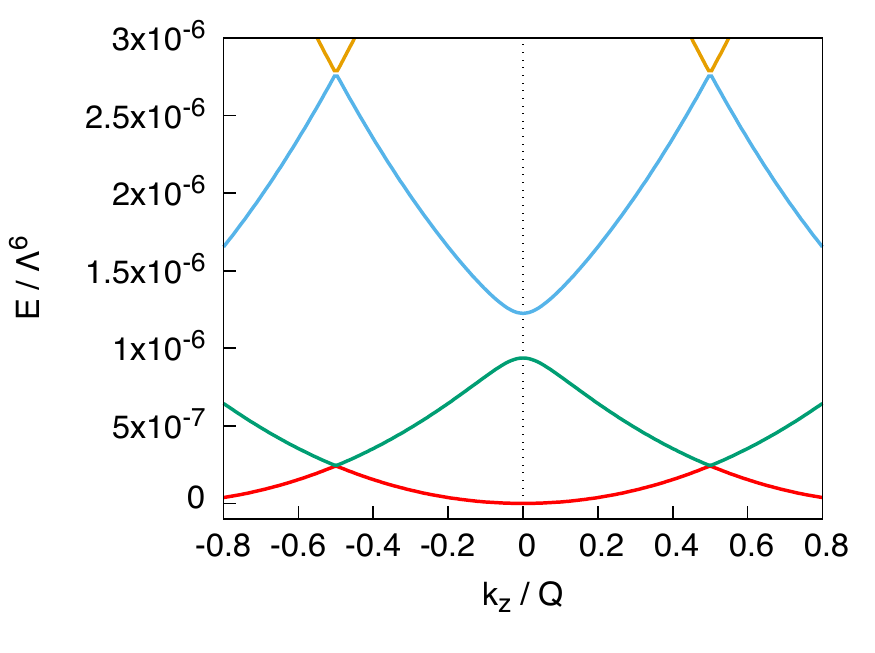}
     \includegraphics[width=.49\textwidth]{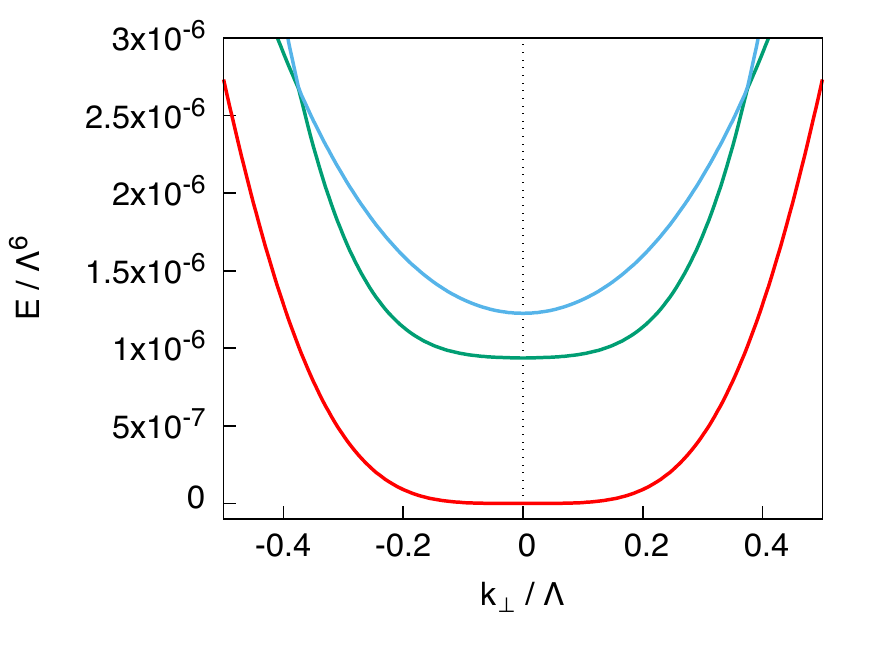}
    \caption{\label{fig:band_structure}
    Eigenvalues of $H_u$ in \eqref{eq:Hu} at $(T,\mu) = (70 , 286.0)$ [MeV],  
    for $k_\bot=0$ ({\bf top}) and for $k_z=0$ ({\bf bottom}). 
    The domain $-0.5\leq k_z/Q \leq 0.5$ is the first Brillouin zone. 
    $k_z$ and $E$ are normalized by $Q$, and $k_{\perp}$ is
    normalized by the UV cutoff scale $\Lambda$.  
    }
   \end{figure}
   
   In Fig.~\ref{fig:band_structure}, we show the eigenvalue 
   $E$ numerically computed for varying $k_z$ and $k_{\perp}$. 
   A marked difference from the eigenvalue of particles in a free space is that there are 
   \emph{infinitely many} levels for given momenta, 
   in analogy to electrons in metals which develop a band structure. 
  
   It is the lowest eigenvalue $E_0$ (red curves in 
   Fig.~\ref{fig:band_structure}) that pertains to 
   the free energy of long-wavelength phonons. 
   By adopting a variational approach, one can rigorously show 
   that $E_0$ behaves for $k_z \sim k_\bot\sim0$ as
   \ba
     E_0 \sim B k_z^2 + C k_{\perp}^4 \,,
     \label{eq:dispersion_phonon}
   \ea
   where $B$ and $C$ are functions of $T$ and $\mu$. 
   The absence of the $\calO(k^2_{\perp})$ term in \eqref{eq:dispersion_phonon} 
   is guaranteed by the property \eqref{eq:ava_g_1}. The proof of 
   \eqref{eq:dispersion_phonon} is somewhat technical and is relegated to Appendix 
   \ref{sec:solut-schl-equat}. 
   Equation \eqref{eq:dispersion_phonon} shows that the elastic free energy 
   of low-energy phonons becomes  
   \ba
     F^{u}_{\rm el} = \frac{1}{2}\int \dd^3x 
     \kkakko{B(\der_z u)^2 + C(\nabla^2_{\perp} u)^2 }.
   \ea
   
   One may suspect that the coefficient $B$ would be given 
   by $\oint f_1$. However, as shown in Appendix~\ref{sec:solut-schl-equat}, 
   this naive guess is incorrect; the coupling between $\phi_0$ and 
   $\phi_{n\ne 0}$ is not negligible even in the perturbation series in $k_z$. 

   \begin{figure}[t]
     \centering 
     \includegraphics[width=.49\textwidth]{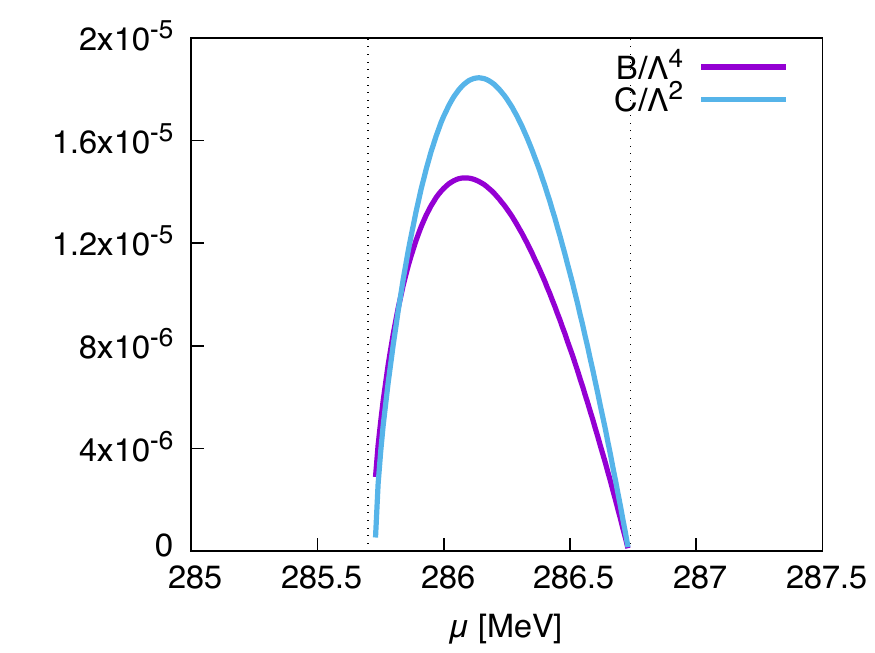}
     \includegraphics[width=.49\textwidth]{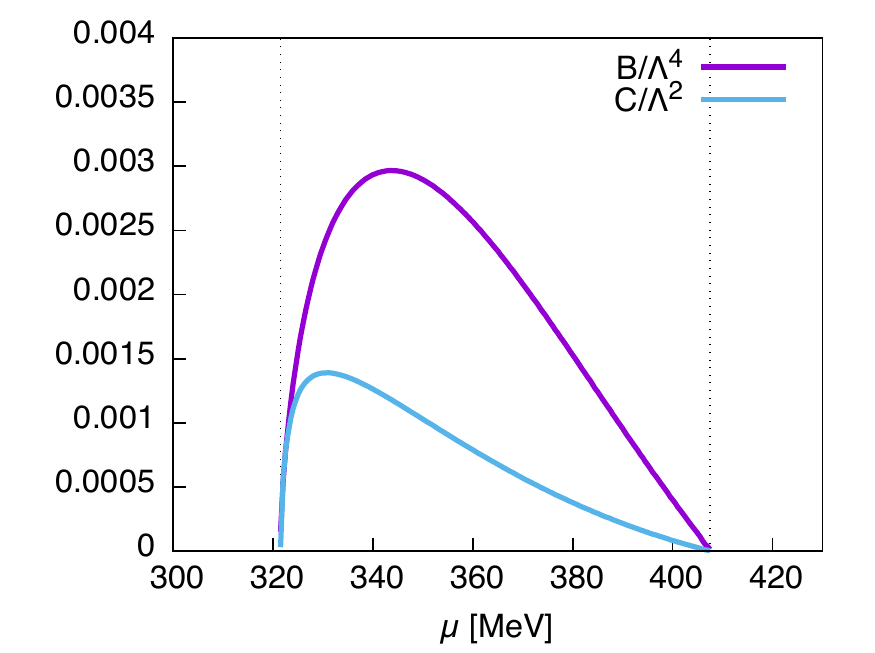}
     \caption{\label{fig:coefs} 
     $B$ and $C$ at $T = 70$ MeV ({\bf top}) and $T = 10$ MeV
     ({\bf bottom}). $B$ and $C$ are normalized by the UV cutoff parameter
     $\Lambda$ with appropriate dimensions. 
     The dotted vertical lines mark the boundaries of the modulated phase.
     }  
   \end{figure}
   
   To extract the values of $B$ and $C$ from eigenvalues, 
   we have numerically fitted
   the curve of $E_0$ with trial functions $E_0= B k_z^2$ and $E_0 =
   C k_{\perp}^4$, respectively, with $B$ and $C$ treated as fitting parameters. 
   Figure \ref{fig:coefs} shows $B$ and $C$ at $T=70$ MeV and $T=10$
   MeV obtained this way. Notably, $B$ and $C$ in Fig.~\ref{fig:coefs} are 
   positive throughout the real kink crystal phase, which proves 
   local stability of this condensate in agreement with numerical 
   results in \cite{Abuki:2011pf}. 
   Because the phonon mode exists only 
   in the real kink crystal phase, it is natural that 
   both coefficients tend to zero at the 
   phase boundaries, although the eigenvalues $E$ were 
   too small near the left boundary for $T=70$ MeV 
   to perform a reliable fitting.  
   
   We mention that the spectrum of gapless excitations over the 
   same background \eqref{eq:Mzero} has also been worked out 
   in \cite{Takahashi2013x} with entirely different methods. 
   However a direct comparison is difficult owing to the fact that 
   the model in \cite{Takahashi2013x} 
   is non-relativistic and in one space dimension, 
   while our model is relativistic and in three space dimensions.

   \subsection{IR divergence and quasi-long-range order}
   \label{sec:IRdiv}
   
   Next, we wish to evaluate the impact of thermal fluctuations 
   of phonons on the stability of the real kink crystal. 
   Taking the Fourier decomposition 
   $M(\bm{x}) = \sum_{n} \M_n \ee^{i n Q (z + u(\bm{x}))}/ \sqrt{L}$, 
   treating $u$ in the Gaussian approximation and ignoring the pion 
   fluctuation, we obtain
   \ba
     \begin{split}
     \langle M(\bm{x}) \rangle & = 
     \frac{1}{\sqrt{L}} \sum_{n} \M_n \langle \ee^{i n Q (z + u(\bm{x}))} \rangle
     \\
     & = \frac{1}{\sqrt{L}} \sum_{n} \M_n \ee^{inQz} 
     \exp\Big[ \! -\frac{1}{2}n^2Q^2\langle u^2\rangle \Big] 
    \end{split}
    \label{eq:Mvev}
    \ea
    with
    \ba
    \begin{split}
     \langle u^2 \rangle & = \frac{ 2\pi}{(2\pi)^3} 
     \int_{\ell_{\perp}^{-1}}^{\Lambda} \dd k_{\perp}\,k_{\perp}
     \int_{-\Lambda}^{\Lambda} \dd k_{z} 
     \frac{T}{ B k_{z}^2 + C k_{\perp}^4}
     \\
     &\sim \frac{T}{4\pi \sqrt{BC}} \log \frac{\ell_{\perp}}{\sqrt{C/B}} \;,
     \end{split}
     \label{eq:exp_phonon}
   \ea
   where we have only incorporated the lowest Matsubara mode since 
   it is dominant in the infrared. The momentum integral is 
   IR divergent and is regularized by a cutoff $\ell_\bot$.  
   One can regard $\ell_{\perp}$ as the transverse 
   diameter of the quark matter in a compact star. 
   In the thermodynamic limit $\ell_\bot\to \infty$, 
   the condensate \eqref{eq:Mvev} drops to zero with negative powers 
   of $\ell_\bot$, implying that the one-dimensional modulation is 
   wiped out by thermal fluctuations at any low $T>0$, a phenomenon 
   known as the Landau-Peierls instability.  
   In fact, if the average amplitude of displacement fluctuation exceeds 
   the interval of layers, it does not make sense to speak of 
   a spatial long-range order.    
   We emphasize that this instability persists even at nonzero 
   quark masses, since it originates from phonons that remain elastic 
   regardless of the quark masses.
   
   \begin{figure}[t]
     \centering
     \includegraphics[width=.49\textwidth]{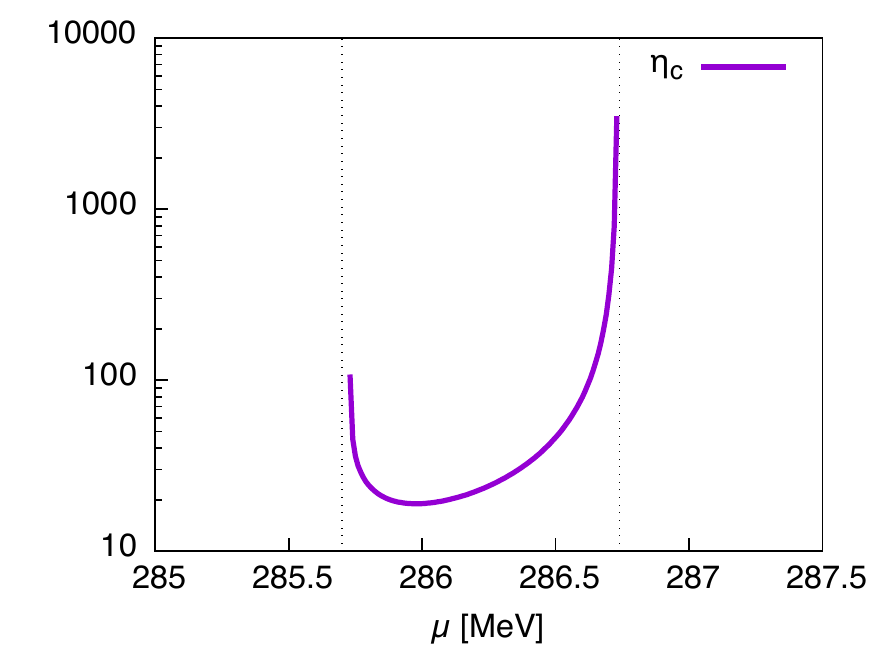}
     \includegraphics[width=.49\textwidth]{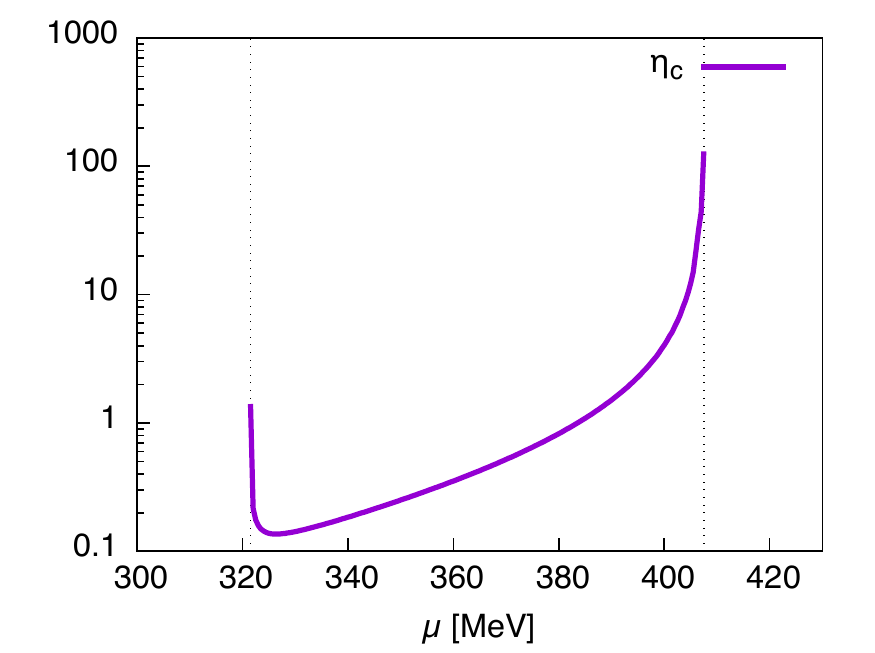}
     \caption{\label{fig:exponent_eta} 
       The exponent $\eta_c$ characterizing the algebraic decay 
       of the order parameter correlation function in the
       kink crystal phase at $T = 70$ MeV ({\bf top}) and $T=10$ MeV 
       ({\bf bottom}). 
       The dotted vertical lines mark the boundaries of the modulated phase.
     }  
   \end{figure}
   In the thermodynamic limit, the system instead exhibits 
   a \emph{quasi-long-range order}. Expanding $M$ in Fourier series and  
   ignoring non-Gaussian effects and pion fluctuations,  
   we find that the correlation function of the order parameter 
   behaves for $|\bm{x}| \gg L$ as 
   \ba
    & \langle M(\bm{x}) M(0)\rangle 
    \notag
    \\
    & = \sum_{n,m} \frac{\M_n \M_m}{L}
    \ee^{i n Q z} \left\langle \exp
    \kkakko{i Q \mkakko{n u(\bm{x}) + m u(\bm{0})}}\right\rangle
    \notag
    \\
    & = \sum_{n,m}
    \frac{\M_n \M_m}{L} \ee^{i n Q z}\exp\kkakko{-
    \frac{Q^2}{2}\left\langle(n u(\bm{x}) + m u(\bm{0}))^2\right\rangle}
    \notag
    \\
    & = \scalebox{0.9}{$\displaystyle 
    \sum_{n,m} \frac{\M_n \M_m}{L} \ee^{i n Qz}
    \exp \! \kkakko{-
    \frac{Q^2}{2}T \! \! \int \!\! \frac{\dd^3 k}{(2 \pi)^3} \frac{m^2+n^2 + 2 m n
    \ee^{i \bm{k\cdot x}}}{Bk_{z}^2 + C k^4_{\perp}}}
    $}
    \notag
    \\
    & = \scalebox{0.9}{$\displaystyle 
    \sum_{n,m} \frac{\M_n \M_m}{L} \ee^{i nQ z}
    \delta_{n,-m} \,
    \exp \! \kkakko{- n^2 Q^2 T \!\! 
    \int \!\! \frac{\dd^3 k}{(2 \pi)^3}
    \frac{1 -  \ee^{i\bm{k\cdot x}}}{Bk_{z}^2 + C k^4_{\perp}}}
    $}
    \notag
    \\
    & \approx  
    \sum_{n\geq 1}  \frac{|\M_n|^2}{L}\left\{
    \!\! 
    \begin{array}{ll}
      \;2 \cos(nQz) \times |z|^{-n^2 \eta_c} & (\bm{x}_{\perp}=\bm{0})
      \\
      \; |x_{\perp}|^{-2 n^2 \eta_c} & (z=0)
    \end{array} . 
    \right.   
    \label{eq:powerdecay}
   \ea
   In the intermediate step, we have dropped terms with 
   $n\ne -m$ since their momentum integrals are infrared divergent. 
    The exponent $\eta_c$ above is defined by 
    \ba
      \eta_c = \frac{Q^2 T}{8 \pi \sqrt{BC}}\;,
      \label{eq:def_of_eta}
    \ea
    which was originally introduced by Caill\'{e} \cite{Caille1972}. 
    Such an algebraic decay of the order parameter correlation 
    has been known to appear in 
    the smectic-A phase of liquid crystals 
    \cite{landau1969,deGennesbook,Chaikin2000,deJeu:2003zz},  
    in the FFLO phases of fermionic superfluids 
    \cite{radzihovsky2008,radzihovsky2011a,radzihovsky2011b} 
    and in the modulated pion condensation in nuclear 
    matter \cite{Baym:1982ca}. Such a slow decay of the order 
    parameter suggests that it would be hard in practice to distinguish 
    it from a true long-range order.  
    
    In Fig.~\ref{fig:exponent_eta}, we show the critical exponent
    $\eta_c$ at $T = 70$ MeV and $T = 10$ MeV. Within numerical 
    precision, $\eta_c$ appears to go to infinity at the ends of the 
    modulated phase. This is physically acceptable because 
    an \emph{exponential} decay of the connected two-point 
    function is expected both in the symmetric phase and 
    in the homogeneous broken phase. 
    
    Although $\langle M(\bm{x})\rangle=0$, 
    a closer look at steps leading to \eqref{eq:powerdecay} 
    shows that $\langle M(\bm{x})^2 \rangle\ne 0$ in this phase. 
    The real kink crystal phase is thus characterized by 
    a homogeneous \emph{higher-order condensate} consisting of 
    four quarks, which is qualitatively distinct from 
    the naive mean-field phase with $\langle M(\bm{x}) \rangle\ne 0$. 
    These phases may be distinguished by 
    $(\ZZ_2)_R\times(\ZZ_2)_L$ symmetry that flips signs of quarks 
    of each chirality independently. 
    Such a novel higher-order condensate is discussed 
    for the FFLO phase of nonrelativistic fermions in 
    \cite{radzihovsky2008,radzihovsky2011a,radzihovsky2011b}.  
    
   The large fluctuation of phonons can be suppressed by several
   factors: (I) strictly zero temperature, (II) higher-dimensional
   modulations, (III) coupling to an external vector
   field (e.g., a magnetic field) and (IV) a finite volume. 
   In case II, the phonon fluctuations no longer cause 
   infrared singularity because the number of spatial directions 
   with a quadratic dispersion decreases \cite{landau1969}. 
   However, it is still an open problem whether higher-dimensional modulations 
   can be energetically favored over a quasi-long-range-ordered phase. 
   Case III is due to the fact that, as laid out in Appendix
   \ref{app:generaleft}, the explicit breaking of rotational invariance
   leads to a non-vanishing $(\nabla_\bot u)^2$ term. This leads to an
   interesting observation that inhomogeneous condensates in QCD under
   magnetic fields \cite{Son:2007ny,Basar:2010zd,Frolov:2010wn,
   Eto:2012qd,Tatsumi:2014wka} could be stable against fluctuations.
   
   Next, we shall analyze the finite-volume effect of case IV in detail.  
   If the system size is finite, $\langle u^2 \rangle$ remains
   finite because the IR divergence is cut off.  
   This is true for putative quark matter inside neutron stars. 
   Roughly speaking, 
   if $\langle u^2\rangle < L^2$, the one-dimensional structure 
   is expected to remain, while it is likely to be wiped out when 
   $\langle u^2\rangle > L^2$. Using \eqref{eq:exp_phonon}, 
   we define the crossover length $\xi_{\perp}$ as a scale 
   below which the one-dimensional modulation persists, 
   \ba
     \xi_{\perp} & = \sqrt{C/B} \;\ee^{4\pi L^2 \sqrt{BC}/T}\,, 
   \ea
   which grows rapidly as the temperature decreases. 
   The periodic structure of the condensate will survive if the 
   size $\ell$ of the quark matter is in the window 
   $L \ll \ell \lesssim \xi_\bot$. 
   
   \begin{figure}[t]
     \centering 
     \includegraphics[width=.49\textwidth]{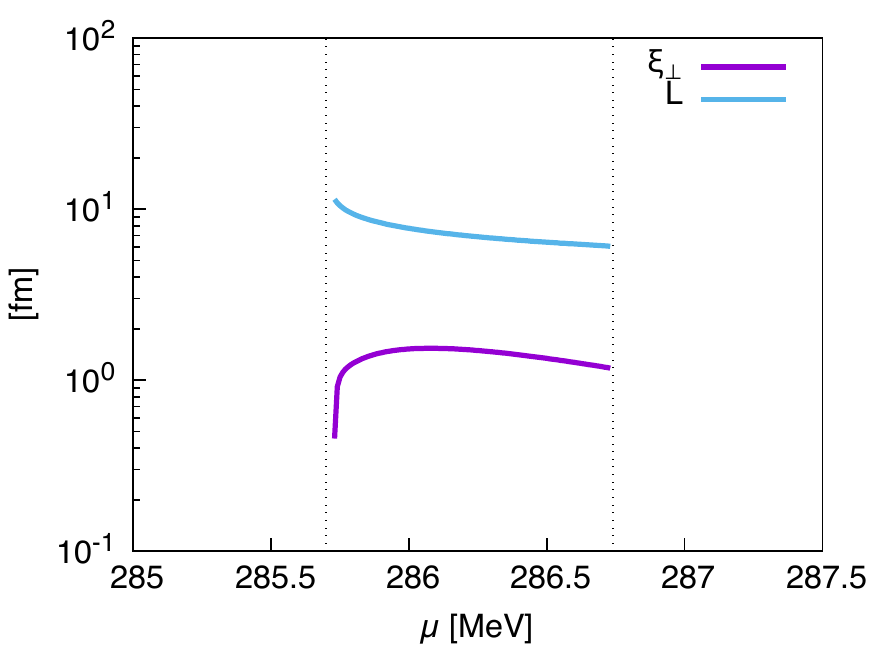}
     \includegraphics[width=.49\textwidth]{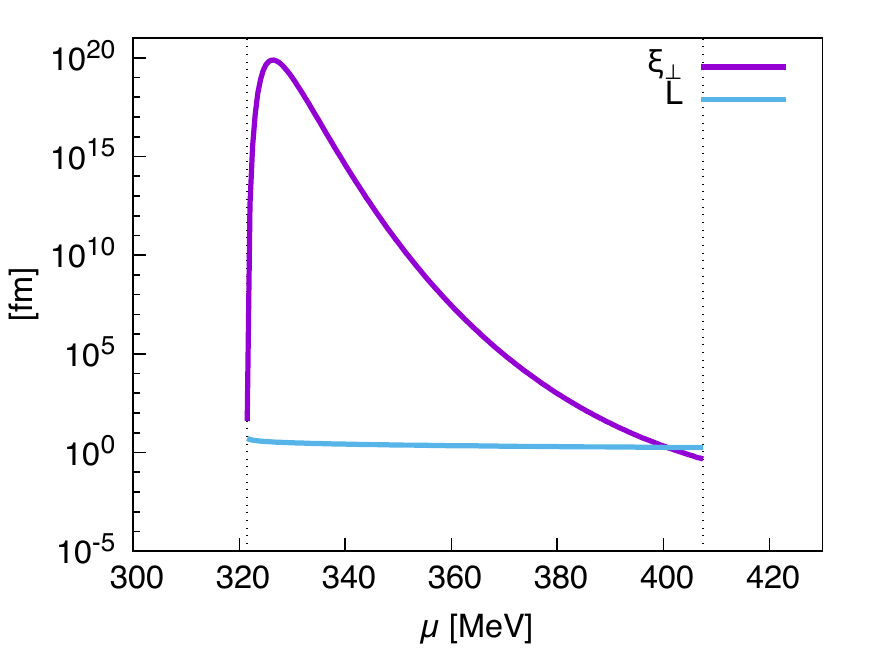}
     \caption{\label{fig:corssover_scale} 
       The crossover scale $\xi_\bot$ and the period $L$ 
       of the kink crystal at $T = 70$ MeV ({\bf top}) 
       and $T=10$ MeV ({\bf bottom}). 
       The dotted vertical lines mark the boundaries of the modulated phase.
       Although $L\equiv 2\pi/Q$ diverges at the left boundary of the modulated phase 
       (recall Fig.~\ref{fig:order_parameters}), 
       this is not visible in these figures due to limited numerical precision.
     }  
   \end{figure}
   Figure~\ref{fig:corssover_scale} shows the 
   crossover scale $\xi_\bot$ together with the period $L$ 
   of the real kink crystal at $T=70$ MeV and $T=10$ MeV. 
   It is observed that $L$ is of order $1$\,fm 
   throughout the modulated phase, while $\xi_\bot$ has a strong 
   $T$ and $\mu$-dependence.  
   At high $T$, soft phonons are so easily excitable that 
   the crossover length is comparable with the period. Thus, there 
   is no vestige of a one-dimensional crystal at high temperature. 
   On the other hand, at low $T$ ($\lesssim 10$ MeV), the situation is different. 
   As shown in the lower panel of Fig.~\ref{fig:corssover_scale}, there is a
   region in which $\xi_{\perp}$ is macroscopic, i.e., of order 
   1\,km ($=10^{18}$\,fm), which would be presumably large enough 
   to accommodate a quark core of neutron stars. Although the 
   GL expansion is not quantitatively reliable at such a low temperature, 
   our conclusion that the real kink crystal condensate persists only 
   at very low temperatures should be qualitatively correct.

   \subsection{Pions}
   \label{sec:pions-kink-crystal}
   
   Next we proceed to the analysis of gapless pions that stem from the
   spontaneous breaking of $\SU(2)_R \times \SU(2)_L$ to
   $\SU(2)_V$. Unlike phonons that only emerge in the modulated 
   phase, pions show up both in the homogeneous broken phase and 
   in the real kink crystal phase, and hence it is of great interest to investigate 
   the nature of pion fluctuations across the transition between these 
   two phases.  Since the vectorial isospin symmetry is intact in both 
   phases we will only take account of the neutral pion fluctuation for 
   simplicity, and defer a fuller analysis to Appendix \ref{sec:gl-expansion-with}.  
   As in \cite{Nickel:2009ke,Nickel:2009wj}, let us make $M$ complex as  
   $M(\bm{x})\equiv -2G[S(\bm{x})+iP_3(\bm{x})]$, with  
   $S(\bm{x})=\langle \bar\psi\psi(\bm{x}) \rangle$ and 
   $P_3(\bm{x})=\langle\bar\psi i\gamma_5 \tau^3\psi(\bm{x})\rangle$.  
   It has been shown by Nickel \cite{Nickel:2009ke,Nickel:2009wj}
   that, assuming a one-dimensional modulation, the $(3+1)$-dimensional
   GL Lagrangian for $M$ can be obtained from that of the chiral 
   Gross--Neveu model \cite{Basar:2008im,Basar:2008ki,Basar:2009fg}, leading to the result
   \ba
    \begin{split}
    \Omega_{\rm GL}[M(\bm{x})] & = \alpha_2 |M|^2 + \alpha_4 \ckakko{|M|^4+|\nabla M|^2} 
    \\
    & \quad  
    + \alpha_6 \big\{ 2|M|^6 + 8|M|^2|\nabla M|^2 
    \\
    & \quad +2\Re\kkakko{(\nabla M)^2 M^{*2}}
    +|\Delta M|^2 \big\} \,. 
    \end{split}
    \label{eq:glbase}
   \ea
   Now we shall follow the same route as for phonons. 
   In the following, $\pi_0$ will be denoted as $\pi$ 
   in order not to clutter notation.  Substituting
   \ba
     M(\bm{x}) & = M_0(z) \ee^{i \pi (\bm{x})}
   \ea
   into (\ref{eq:GL_NJL4}) and expanding in $\pi$, we obtain
    \ba
    \begin{split}
      & \Omega_{\rm GL}[M(\bm{x})]   
      \\
      & = \Omega_{\rm GL}[M_0(z)] +  \frac{f_{1\pi}(z)}{2}(\der_z  \pi)^2
      +\frac{f_{2\pi}(z)}{2}(\der_z^2  \pi)^2 
      \\
      & \quad + \frac{g_{1\pi}(z)}{2}(\nabla_\perp \pi)^2 
      + \frac{g_{2\pi}(z)}{2}(\nabla_{\perp}^2\pi)^2 
      \\
      & \quad 
      + h_{1\pi}(z) (\der_z \pi)(\nabla_{\perp}^2 \pi) 
      \\ 
      & \quad + {h_{2\pi}}(z) (\der_z^2 \pi)
      (\nabla_{\perp}^2\pi) + \calO(\pi^3)\,,
    \end{split}
    \label{eq:OGLeq_pi}
   \ea
   where $f_{1\pi}$, $f_{2\pi}$, $g_{1\pi}$, $g_{2\pi}$,
   $h_{1\pi}$, $h_{2\pi}$ are defined as
   \ba
     \begin{split}
     f_{1\pi}(z) &= 2 (\alpha_4+6\alpha_6M_0^2) (M_0)^2  
     \\
     & \qquad + 4\alpha_6 \ckakko{2(M_0')^2-M_0M_0''}\,, 
     \\
     f_{2\pi}(z) &= 2 \alpha_6 (M_0)^2\,, 
     \\
     g_{1\pi}(z) &= 2 (\alpha_4+6\alpha_6M_0^2) (M_0)^2 - 4
     \alpha_6 M_0M_0''  \,,
     \\
     g_{2\pi}(z) &=2\alpha_6 (M_0)^2 \;,
     \\
     h_{1\pi}(z) & = 4 \alpha_6    M_0M_0'
     \;,\quad  h_{2\pi}(z) = 2 \alpha_6 (M_0)^2 \;.
     \end{split}
   \ea
   We note that, unlike the phonon,
   \ba
     \oint g_{1\pi} \neq 0\;.
   \ea   
   The eigenvalue $E_\pi$ of pions may be derived from 
   the eigenvalue equation
   \ba
     E_{\pi} \pi = \frac{\delta F[\pi]}{\delta \pi}
   \ea 
   with
   \ba
     \begin{split}
     F[\pi] & \equiv \int \dd^3x ~\Bigl\{
     \frac{f_{1\pi}(z)}{2}(\der_z  \pi)^2
     +\frac{f_{2\pi}(z)}{2}(\der_z^2 \pi)^2 
     \\
     & \quad  
     + \frac{g_{1\pi}(z)}{2} (\nabla_\perp \pi)^2 + 
     \frac{g_{2\pi}(z)}{2}(\nabla_{\perp}^2\pi)^2
     \\
     & \quad  
     + h_{1\pi}(z) (\der_z \pi)(\nabla_{\perp}^2 \pi) 
     + h_{2\pi}(z) (\der_z^2 \pi)(\nabla_{\perp}^2 \pi) \Bigr\} \,.
     \end{split}
     \hspace{-10pt}
   \ea 
   In a more explicit form, it reads 
   \ba
     \begin{split}
    H_\pi \pi = E_{\pi} \pi 
     \end{split}
     \label{eq:schrodinger_eq_pi}
   \ea
   with
   \ba
   \begin{split}
    H_\pi &\equiv - \partial_z (f_{1\pi} \partial_z)+\partial_z^2 (f_{2\pi}
     \partial_z^2) - g_{1\pi} \nabla_{\perp}^2 + g_{2\pi} \nabla_{\perp}^4   
     \\
     &\quad- (\der_z h_{1\pi}) \nabla_{\perp}^2 
     + \nabla_{\perp}^2\ckakko{h_{2\pi},\der_z^2}_+ \,. \label{eq:Hpi}
     \end{split}
   \ea
   Since the structure of \eqref{eq:schrodinger_eq_pi} is identical to the 
   case of phonons \eqref{eq:schrodinger_eq}, one can apply 
   the same techniques to solve it. On the basis of Bloch's theorem, 
   $\pi$ may be decomposed as
   \ba
     \pi(\bm{x}) =  \ee^{i\bm{k}_{\perp}\cdot \,
     \bm{x}_{\perp}}
     \ee^{ i k_z z} \phi(z) \,,
   \ea
   for a crystal momentum $k_z$ and transverse momenta $k_\bot$. 
   $\phi(z)$ is a periodic function with period $L$. Substituting this 
   decomposition, we arrive at an eigenvalue equation for $\phi$, which 
   can be solved with the same numerical methods as for phonons.
   
   \begin{figure}[t]
     \centering 
     \includegraphics[width=.49\textwidth]{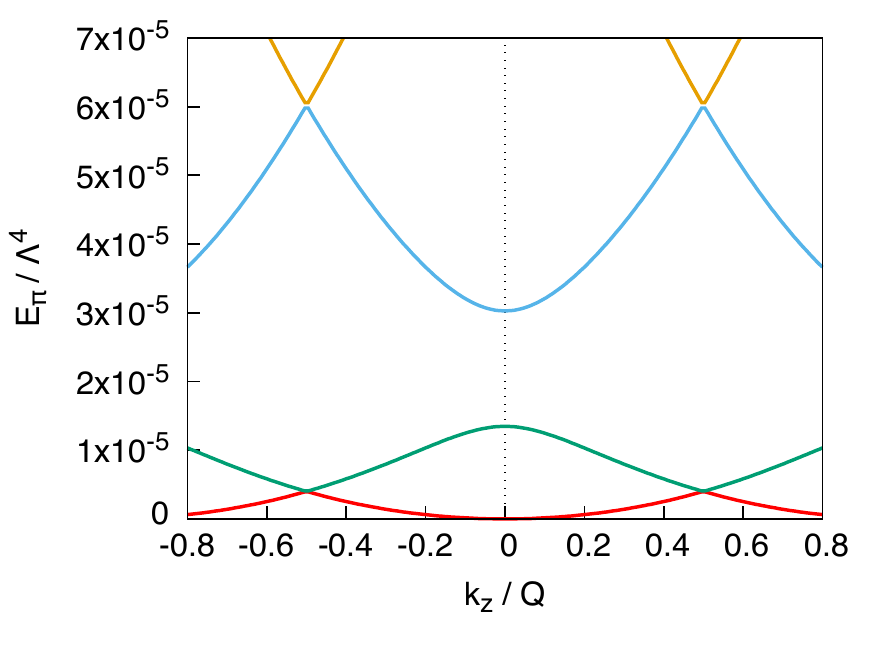}
     \includegraphics[width=.49\textwidth]{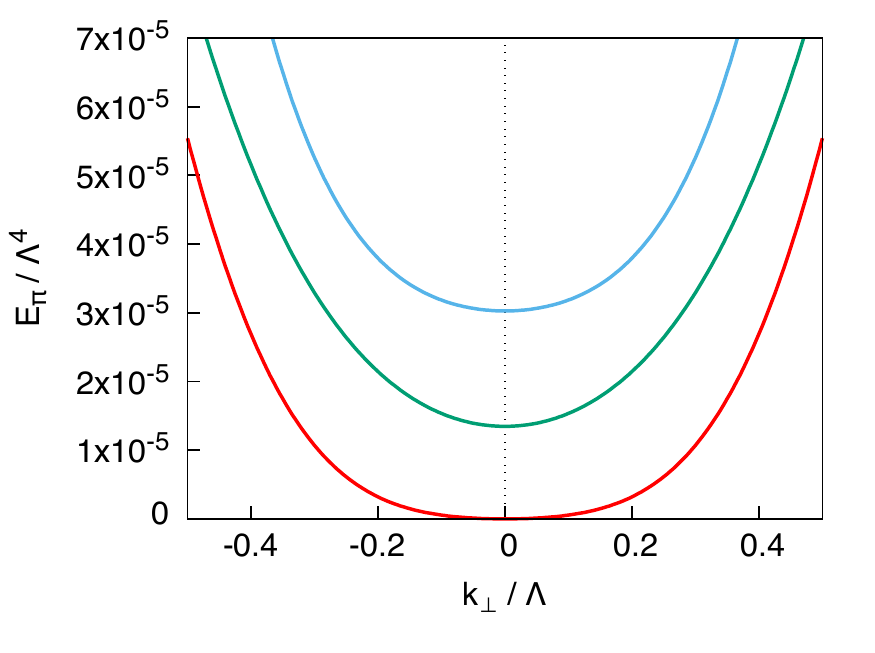}
     \caption{\label{fig:band_structure_pi} 
     Eigenvalues of $H_\pi$ in \eqref{eq:Hpi} at $(T,\mu) = (70 , 286.0)$
     [MeV], for $k_\bot=0$ ({\bf top}) and for $k_z=0$ ({\bf bottom}). 
     The domain $-0.5\leq k_z/Q \leq 0.5$ is the first Brillouin zone. 
     $k_z$ and $k_\bot$ are normalized by $Q$ and $\Lambda$, respectively. 
     }  
   \end{figure}
   In Fig.~\ref{fig:band_structure_pi}, we show the eigenvalues 
   $E_{\pi}$ for varying $k_z$ and $k_{\bot}$.  To analyze the lowest 
   eigenvalue $E_{\pi,0}$ (red curves in Fig.~\ref{fig:band_structure_pi}), 
   we have used a variational method along the lines of 
   Appendix \ref{sec:solut-schl-equat}, which shows that 
   the leading behavior of $E_{\pi,0}$ near $k_{z}=k_{\bot}=0$ 
   is given by 
   \ba
     E_{\pi,0} \sim F^2_{\parallel} k_{z}^2 + F^2_{\perp} k_{\perp}^2\,,
   \ea
   where
   \ba
     F^2_{\perp} \equiv  \oint g_{1\pi}
     \label{eq:3}
   \ea
   and $F^2_{\parallel}$ is a positive function that has to be 
   computed numerically. 
   The elastic free energy of low-energy pions is therefore
   \ba
      F^{\pi}_{\rm el} = \frac{1}{2} \int \dd^3x 
      \kkakko{F_{\parallel}^2(\der_z \pi)^2 +
      F_{\perp}^2(\nabla_{\perp} \pi)^2 }\;. 
   \ea
   Since $F_\bot^2\ne0$, pions on the real kink crystal 
   have a linear dispersion in all directions, in contrast to phonons. 
   Thus the thermal fluctuations of pions neither cause any
   infrared divergence nor destroy the long-range order.
   
   \begin{figure}[t]
     \centering 
     \includegraphics[width=.49\textwidth]{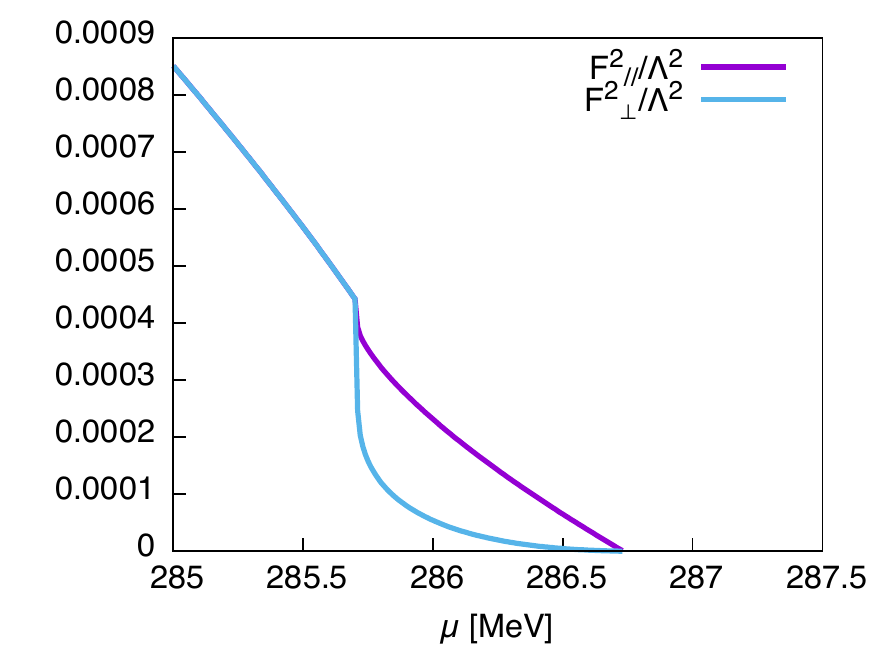}
     \includegraphics[width=.49\textwidth]{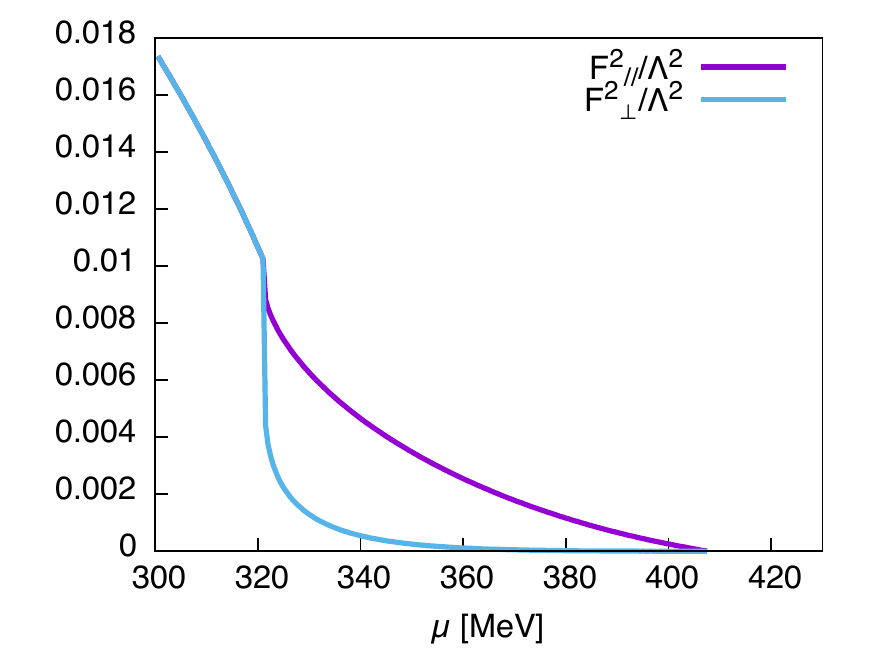}
     \caption{\label{fig:f_pi} 
     $F_{\perp}^2$ and $F_{\parallel}^2$ at $T = 70$ MeV ({\bf
     top}) and $T = 10$ MeV ({\bf bottom}). 
     }  
   \end{figure}
   
   $F^2_{\perp}$ and $F^2_{\parallel}$ at $T=70$ MeV are shown 
   in Fig.~\ref{fig:f_pi}.   
   In the homogeneous broken phase, the
   system is isotropic and $F^2_{\perp} = F^2_{\parallel}$. In the kink
   crystal phase, the system is anisotropic and the coefficients no longer 
   coincide.  In our calculation, in the kink crystal phase,
   $F^2_{\parallel}> F^2_{\perp}$ holds at any temperature and
   chemical potential.

   \section{Conclusion}
   \label{sec:conclusion}
  
  In this paper, we presented a first systematic study of low-energy fluctuations 
   in the real kink crystal phase of the NJL model. The elastic free energy for
  the phonon was shown to be $(\partial_zu)^2$ in the longitudinal direction 
  and $(\partial_\perp^2u)^2$ in the transverse directions with respect to the 
  modulation of the condensate, 
  which had an important consequence of vanishing order 
  parameter, exhibiting a quasi-long-range order. We argued that, 
  since the correlation function decays only algebraically, the real kink 
  crystal may be sustained in compact star cores if the temperature 
  is sufficiently low. Our analysis, which should be reliable at least near 
  the critical point, suggests that fluctuation effects that are missed 
  in the mean-field treatment change the nature of inhomogeneous 
  chiral condensation qualitatively.   

This work can be extended in various directions. 
The NJL model used here can be extended with the 
inclusion of vector interaction \cite{Carignano:2010ac}, 
or one can use the quark-meson model 
\cite{Nickel:2009wj,Carignano:2014jla}; in either case, 
the phase structure would be quantitatively modified. 
For a fuller understanding of the phonon fluctuation, we 
need to consider anharmonic effects due to the $\calO(u^3)$ 
term in \eqref{eq:uuu}, which is known for the case of smectic liquid crystals 
to modify the scaling behavior \eqref{eq:powerdecay} 
\cite{PhysRevLett.47.856,PhysRevA.26.915}.  
It is also intriguing to study the interaction between quarks, phonons and pions. 
As for the fate of the Lifshitz critical point in Fig.~\ref{fig:phase_withinhomo},  
we point out that it must be eliminated from the QCD phase diagram 
once fluctuations are fully incorporated, because the lower 
critical dimension of the isotropic Lifshitz critical behavior 
with continuous symmetry is 4 \cite{Diehl2002a}. This is true in the 
chiral limit, while for nonzero quark masses, the continuous symmetry 
is broken, and a more careful study is needed. At any rate, the mean-field 
picture can fail even qualitatively near the Lifshitz point, and it is highly  
desirable to develop a method for analyzing fluctuations  
that goes beyond 
the GL expansion. This is a hard problem at this stage but should be 
seriously considered in future research. 
   
Finally we note that for phenomenological applications to the physics of 
compact stars it would be important to incorporate 
effects of nonzero quark masses, isospin chemical potential, 
electromagnetic interactions and color superconductivity, 
which deserves further investigation.  

\vspace{0.5cm}

\noindent\textbf{Note added}
\vspace{0.2cm}\\
While this work was being completed, we learned of an 
independent work \cite{Lee:2015bva} where the  
low-energy fluctuations over a Fulde--Ferrell type inhomogeneous 
chiral condensate were discussed. 
That condensate breaks translational and internal 
symmetries in a different way from the real kink crystal considered 
in our work, and hence produces a different number of 
Nambu--Goldstone modes.

\begin{acknowledgments}
   We thank G.~Baym, T.-G.~Lee and D.~A.~Takahashi for useful discussions. 
   K.\,K.~and T.\,N. were supported by the Special Postdoctoral
   Research Program of RIKEN. 
   Y.\,H. was supported by JSPS KAKENHI Grants No.~24740184.   
   T.\,K. was supported by the RIKEN iTHES project. 
\end{acknowledgments}

\appendix
   
 \section{Phonon effective theory coupled to a vector field}
 \label{app:generaleft}   
  
  In this Appendix we present a simple argument based on the method 
  of \cite{Hidaka:2014fra} showing that 
  a coupling to an external vector field modifies the dispersion 
  of phonons in a qualitative manner. 
  
  Let us start with a theory with no vector field.  
  Suppose $\langle\phi\rangle=\phi_0(z)$ is a modulated static solution that 
  minimizes the rotationally symmetric free energy 
  $F[\phi]=\int \dd^3x~\mathcal{F}(\phi,\der \phi)$. 
  Now we consider a translational fluctuation corresponding to 
  the phonon around this solution, \mbox{$\phi(\bm{x})=\phi_0(z+ u(\bm{x}))$}. 
  Plugging this into $F[\phi]$ and
  expanding in powers of $u$ and $\nabla u$, one obtains the effective
  theory for the $u$ field.
  Since $u$ always appears in the form, $z+u$,
  the effective free energy is constructed from
  \ba
    \label{ingredients}
    \text{scalar functions of $z + u$, ~their derivatives, ~and $\delta_{ij}$}.
  \ea
     More concretely, the ingredients with one and two derivatives are
  \ba
    \quad\partial_i(z+u)&=\delta_i^z+\partial_iu\,,\\
    \partial_i\partial_j(z+u) & =\partial_i\partial_ju\,,
  \ea
  respectively.
  These ingredients are compatible with the original symmetry
  as they should be.  For example, 
  \ba 
    (\delta_i^z+\partial_iu)^2 = 1+2 \der_z u + (\nabla u)^2
  \ea
  automatically reproduces the rotationally invariant combination 
  derived in Sec.~\ref{sec:general-discussion}. 
  It is now straightforward to combine these ingredients and derive 
  the leading phonon free energy 
  \ba
    \mathcal{F} & = 
    A \kkakko{\der_z u + \frac{1}{2} (\nabla u)^2}
    + B \kkakko{\der_z u + \frac{1}{2} (\nabla u)^2 }^2 \,.
    \label{kink_EFT_gapless}
  \ea
  Just as we discussed in Sec.~\ref{sec:general-discussion},
  the total derivative term, $\der_z u$, linear in $u$
  is prohibited by the minimum-energy condition,
  so that we have $A=0$
  and $\mathcal{F}$ reduces to \eqref{eq:uuu} after taking into account higher derivative terms.
  The reader is referred to \cite{Hidaka:2014fra} for more details of 
  this method. 
    
  We shall then proceed to a discussion on the elastic free energy 
  in the presence of an external vector field $v^i$.    
  Such a situation is pertinent to modulated chiral 
  condensates in QCD with an external magnetic field.
  Now one can use $v^i$ in addition to the previous ingredients~\eqref{ingredients},
  so that the following interaction can appear in the free energy for example:
  \ba
    v_i\left(\delta_i^z+\partial_i{u}\right)=v^z+v^i\partial_i{u}\,.
  \ea
  The simplest modification of (\ref{kink_EFT_gapless}) will then be 
  \ba
    \begin{split}
    \mathcal{F} & = 
    A \kkakko{\der_z u + \frac{1}{2} (\nabla u)^2}
    + B \kkakko{\der_z u + \frac{1}{2} (\nabla u)^2 }^2 
    \\
    & \quad    -v^z-v^i\partial_i{u} \,.  
    \end{split}
  \ea
  We notice here that the total derivative terms in the linear order arise
  from both of the $A$ and $v^i$ couplings.  If the vector $v_i$
  is an external field, i.e., a non-dynamical field, and takes a value
  \ba
    v^i = A \delta^i_z\,,
  \ea
  the linear order terms cancel out:
  \ba
   \label{action_external}
    \mathcal{F} = 
    \frac{A}{2} (\nabla u)^2 + B (\der_z u)^2 
    + \calO({u}^3) \,.  
  \ea
  Therefore, when an external field $v^i$ explicitly breaks the rotation symmetry,
  the $(\nabla_\bot{u})^2$ term need not vanish,
  which can be realized without spoiling the minimum-energy condition
  of the condensate. The dispersion of phonons becomes linear in all directions, 
  implying that the severe infrared divergence at finite temperature 
  (cf.~Sec.~\ref{sec:IRdiv}) is ameliorated.  
  
  It would be important to note that the non-vanishing $(\nabla_\bot{u})^2$ term 
  in the free energy does not arise if the vector is \emph{dynamical} 
  (i.e., not external) and its condensation is aligned in the $z$-direction.
  In such a case, 
  we have gapped fluctuations associated with the spontaneously broken rotational symmetries,
  just as in the smectic-A phase of liquid crystals.
  The free energy after integrating out those gapped fluctuations
  turns out to be the same as the one without dynamical vector fields, 
  \eqref{eq:uuu}, which results in a strongly anisotropic dispersion of phonons.

  \section{GL expansion with pions}
  \label{sec:gl-expansion-with}
  
  In this Appendix, we incorporate pionic modes into the effective theory 
  (\ref{eq:glbase}) so that it becomes manifestly invariant under 
  $\SU(2)_R \times\SU(2)_L$. 
  It is convenient to work with the matrix field
  \ba
     \Sigma(\bm{x}) & \equiv -2G[S(\bm{x})\1 + i P_a(\bm{x}) \tau^a] \,, 
     \label{eq:Sigmarep}
  \ea
  where $\{\tau^a\}$ are the Pauli matrices.  Under $U_R\in \SU(2)_{R}$ 
  and $U_L\in \SU(2)_{L}$, it transforms as $\Sigma\to U_L
  \Sigma U_R^\dagger$.  Then the most generic GL function invariant
  under $\SU(2)_{R}\times \SU(2)_{L}$ is given, up to 6th order
  in fields and derivatives, by
   \ba
   \begin{split}
    \Omega_{\rm GL}(\Sigma) & = 
    \frac{\alpha_2}{2}\tr[\Sigma^\dagger\Sigma] 
    + \frac{\alpha_4}{2}\tr[\der_i \Sigma^\dagger \der_i \Sigma]
    + \frac{\alpha_4}{4}\mkakko{\tr[\Sigma^\dagger\Sigma] }^2 
    \\
    & \quad 
    + \frac{\alpha_6}{4}\mkakko{ \tr[\Sigma^\dagger\Sigma] }^3
    + \frac{\alpha_6}{2}\tr[\Delta \Sigma^\dagger \Delta \Sigma] 
    \\
    & \quad + \beta_1 \tr[\Sigma^\dagger\Sigma] \tr [\der_i
    \Sigma^\dagger \der_i \Sigma]
    \\
    & \quad 
    + \beta_2 \big\{ \tr[ (\der_i \Sigma)\Sigma^\dagger (\der_i
    \Sigma)\Sigma^\dagger ] + \text{h.c.} \big\}
    \\
    & \quad + \beta_3 \mkakko{\tr[\Sigma^\dagger \der_i \Sigma]}^2
    + \beta_4 \tr[\Sigma^\dagger \Sigma]\tr[\Sigma^\dagger
    \Delta\Sigma] \,,
    \end{split}
    \label{eq:glexpa}
   \ea
   where the coefficients $\beta_1$, $\beta_2$, $\beta_3$ and $\beta_4$
   are yet to be determined.  However, the last four terms of
   \eqref{eq:glexpa} are not independent, owing to the relations
   \ba
   \begin{split}
    &\tr[ (\der_i \Sigma)\Sigma^\dagger (\der_i \Sigma)\Sigma^\dagger
    ] 
    \\
    &\qquad= \mkakko{\tr[\Sigma^\dagger \der_i \Sigma]}^2 -
    \frac{1}{2}\tr[\Sigma^\dagger \Sigma]\tr[\der_i\Sigma^\dagger
    \der_i \Sigma]\,,
   \end{split}
   \ea
   and
   \ba
     \begin{split}
    &\tr[\Sigma^\dagger \Sigma]\tr[\Sigma^\dagger \Delta\Sigma] 
    \\
    &\qquad = \der_i \mkakko{ \tr[\Sigma^\dagger \Sigma] \tr[\Sigma^\dagger \der_i\Sigma] } 
    - 2 \mkakko{ \tr[\Sigma^\dagger \der_i\Sigma] }^2 
    \\
    &   \qquad\quad
    - \tr[\Sigma^\dagger \Sigma] \tr[\der_i \Sigma^\dagger \der_i\Sigma] \,. 
    \end{split}
   \ea 
   Therefore one can let $\beta_2=\beta_4=0$ without loss of generality
   (up to total derivatives).  Since $\Omega_{\rm GL}(\Sigma)$ must
   reduce to \eqref{eq:glbase} when $P_1=P_2=0$, we require
   \begin{multline*}
     \alpha_6 \ckakko{ 8|M|^2|\nabla M|^2+2\Re\kkakko{(\nabla M)^2 M^{*2}} }
     \overset{!}{=}
     \\
     4 \beta_1 |M|^2|\nabla M|^2 
     + 2 \beta_3 \ckakko{ |M|^2|\nabla M|^2+\Re\kkakko{(\nabla M)^2 M^{*2}} }  \,.
   \end{multline*}
   \ba
     \therefore ~~ \beta_1 = \frac{3}{2}\alpha_6 
     \qquad \text{and} \qquad \beta_3 = \alpha_6 \,.
   \ea
   Thus the desired GL function with manifest chiral symmetry is given by
   \ba
   \begin{split}
    &\Omega_{\rm GL}(\Sigma) 
    \\
    = &  \ 
    \frac{\alpha_2}{2}\tr[\Sigma^\dagger\Sigma] 
    + \frac{\alpha_4}{2}\tr[\der_i \Sigma^\dagger \der_i \Sigma]
    + \frac{\alpha_4}{4}\mkakko{\tr[\Sigma^\dagger\Sigma] }^2 
    \\
    & + \frac{\alpha_6}{4}\mkakko{ \tr[\Sigma^\dagger\Sigma] }^3
    + \frac{\alpha_6}{2}\tr[\Delta \Sigma^\dagger \Delta \Sigma] 
    \\
    &+ \frac{3\alpha_6}{2} \tr[\Sigma^\dagger\Sigma] 
    \tr [\der_i \Sigma^\dagger \der_i \Sigma]
    + \alpha_6 \mkakko{\tr[\Sigma^\dagger \der_i \Sigma]}^2 .
    \end{split}
    \label{eq:GL_su2xsu2}
   \ea

\section{Regularization of the thermodynamic potential}
\label{app:propertime}

   In the main body of this paper, we have employed 
   the momentum cutoff to regularize UV divergences in the NJL model. 
   Although the momentum cutoff is subtle in the presence of a modulated 
   condensate as is claimed in \cite{Nickel:2009ke,Buballa:2014tba}, 
   our work is concerned with the GL expansion around a homogeneous   
   vacuum, and so this problem is unlikely to obstruct our present analysis. 
   We would also like to add that the phase structure including a modulated phase has 
   been found to be quite robust against varying regularizations \cite{Partyka:2008sv}.  
   That being said, however, it would be desirable for theoretical completeness 
   to have an alternative derivation of regularized GL coefficients  
   that is free from the above subtlety. In this Appendix we shall 
   demonstrate one of such regularization methods.

Let us start from the mean-field thermodynamic potential 
of the NJL model in Euclidean spacetime, 
\ba
  \begin{split}
  \Omega_{\rm MF}(T,\mu) & = - N_cN_f \log   
  \det\kkakko{\slashed{\der}-\mu\gamma_4+M(\bm{x})} 
  \\
  & \qquad + \int \dd^4x~\frac{M(\bm{x})^2}{4G}\,. 
  \end{split}
\ea
The functional determinant suffers from UV divergences, and one has to specify a regularization scheme. 
The conventional three-momentum cutoff is not useful  
for a generic inhomogeneous mean field, whereas 
the Schwinger proper-time regularization \cite{Schwinger:1951nm} 
is tricky at nonzero chemical potential \cite{Inagaki:2003ac,Inagaki:2015lma}. 
Now we introduce a new regularization scheme free from these problems, based on 
the formula 
\ba
  \log A - \log B & = - 
  \int_0^\infty \!\!\! \dd w w \ckakko{
    \frac{1}{(w+A)^2} - \frac{1}{(w+B)^2}
  }\,.  
\ea 
See \cite{Litim:2001ky,Litim:2002xm} 
for the relation of this approach to the naive proper-time regularization. 
The methodology of derivative expansion is well known \cite{Eguchi:1976iz}, 
and we shall be brief here, 
\begin{widetext}
\begin{align*}
   & \log \frac{
     \det [\slashed{\der}-\mu\gamma_4+M] 
   }{
     \det [\slashed{\der}-\mu\gamma_4]
   }
   = \frac{1}{2}\Tr \ckakko{
   \log \kkakko{
     - (\der_4-\mu)^2 - \Delta + (\slashed{\nabla}M) + M^2
   } - \log \kkakko{
     - (\der_4-\mu)^2 - \Delta 
   } }
   \\
   & = - \frac{1}{2} \tr \int_x \int_p 
   \int_0^{\Lambda^2} \dd w w \ee^{-ipx} \ckakko{
     \frac{1}{[w - (\der_4-\mu)^2 - \Delta + (\slashed{\nabla}M) + M^2]^2} 
     - \frac{1}{[w - (\der_4-\mu)^2 - \Delta]^2}
   }
  \ee^{ipx}
  \\
  & = - \frac{1}{2} \tr \int_x \int_p 
  \int_0^{\Lambda^2} \dd w w \ckakko{
     \frac{1}{[w + (p_4 + i\mu)^2 - (\nabla+i\pp)^2 + (\slashed{\nabla}M) + M^2]^2} 
     - \frac{1}{[w + (p_4 + i\mu)^2 + \pp^2]^2}
  }
  \\
  & = \frac{1}{2} \int_x \int_p 
   \int_0^{\Lambda^2} \dd w \ckakko{
     \frac{2w}{(w+P^2)^3} \tr \hat O - \frac{3w}{(w+P^2)^4} \tr \hat O^2 
     + \frac{4w}{(w+P^2)^5} \tr \hat O^3 - \frac{5w}{(w+P^2)^6} \tr \hat O^4 + \dots
   } \,,
\end{align*}
\end{widetext}
where the ``$\tr$'' denotes a trace over spinor indices, 
$\Lambda$ is a UV cutoff, $\displaystyle \int_x \equiv \int \dd^4x$,  
$\displaystyle \int_p \equiv T\sum_{p_4}\int \frac{\dd^3 p}{(2\pi)^3}$,  
$\hat O \equiv -2i\pp\cdot \nabla - \Delta + (\slashed{\nabla}M) + M^2$  
and $P^2 \equiv (p_4+i\mu)^2+\pp^2$.  For the purpose of obtaining 
the GL expansion up to 6th order, it suffices to expand only up to $\hat O^4$. 
After a bit of algebra involving a trick $p_i p_j \to \delta_{ij}\pp^2/3$ and 
integration by parts, we obtain 
\ba
  \begin{split}
  \tr \hat O & = 4M^2\,,
  \\
  \tr \hat O^2 & = 4[(\nabla M)^2+M^4] \,,
  \\
  \tr \hat O^3 & = 4[M^6+7M^2(\nabla M)^2+(\Delta M)^2] \,,
  \\
  \tr \hat O^4 & = \frac{16}{3}\pp^2[4M^2 (\nabla M)^2+(\Delta M)^2] \,,
  \end{split}
\ea
where we have omitted total derivatives, terms that are higher order in the GL expansion, 
and terms odd in $\pp$.  Using a formula 
$\int_p \frac{\pp^2}{(w+P^2)^6} = \frac{3}{10}\int_p \frac{1}{(w+P^2)^5}$, 
we can match the expanded determinant with \eqref{eq:GL_NJL4} 
and extract the regularized GL coefficients as
\begin{subequations}
\ba
  \alpha_2 & = \frac{1}{4G} - 4 N_cN_f \int_p \int_0^{\Lambda^2} \dd w \frac{w}{(w+P^2)^3} \,, 
  \\
  \alpha_4 & = 6 N_cN_f \int_p \int_0^{\Lambda^2} \dd w \frac{w}{(w+P^2)^4} \,, 
  \\ 
  \alpha_6 & = - 4 N_cN_f \int_p \int_0^{\Lambda^2} \dd w \frac{w}{(w+P^2)^5} \,. 
\ea
\label{eq:alphafinite}%
\end{subequations}
In the limit $\Lambda\to \infty$, we formally recover \eqref{eq:alphas}. 
This derivation, in which the functional determinant is regularized directly, 
satisfies the requirement \cite{Nickel:2009ke,Buballa:2014tba} that surface terms 
in momentum integrals strictly vanish. 
However, the three-momentum cutoff is practically 
more useful and is used throughout the main part of this paper.

   \section{Proof of $\displaystyle \oint g_1 = 0$}
   \label{sec:proof_g1=0}
   
   In this Appendix, we prove \eqref{eq:ava_g_1}, i.e., that 
   $g_1(z)$ must vanish in the average sense when $\nu$ and $q$ 
   in \eqref{eq:Mzero} are so tuned that $M_0(z)$ attains the minimum 
   of the free energy per period. To show this, 
   we consider a scaled configuration 
   \ba
     M(\lambda, z) & \equiv M_0(\lambda z)\,,
   \ea
   which has a period $L/\lambda$ with $L$ 
   defined in \eqref{eq:defLQ}. It follows from the 
   minimum-energy requirement for $M_0(z)$ that 
   the GL free energy per period of $M(\lambda,z)$ 
   must have an extremum at $\lambda=1$, namely    
   \ba
     \lim_{\lambda\to 1}\frac{\der}{\der \lambda}
     \kkakko{
       \frac{1}{L/\lambda}\int_0^{L/\lambda}\!\!\!\!\dd z~
       \Omega_{\rm GL}[M(\lambda, z)]
     } = 0\,. 
   \ea
   Note that this equality is trivial if $\Omega_{\rm GL}[M]$ did not 
   depend on the derivatives of $M$, for one can simply get rid of $\lambda$ 
   from inside of $[\dots]$ via a change of variable $z\to z/\lambda$. 
   However, this is not possible for $\Omega_{\rm}[M]$ in \eqref{eq:GL_NJL4} 
   which does depend on the derivatives of $M$. 
   
   An explicit calculation yields
   \ba
    \!\!\!  
    \frac{\der}{\der \lambda}
     \kkakko{
       \frac{1}{L/\lambda}\int_0^{L/\lambda}\!\!\!\!\dd z~
       \Omega_{\rm GL}[M(\lambda, z)]
     } 
    = \fbox{1} + \fbox{2} + \fbox{3}\,, 
    \label{eq:lambdazero}
   \ea
   where [with $M'\equiv \der_z M(\lambda,z)$ and $M''\equiv \der_z^2 M(\lambda,z)$, 
   and omitting the subscript ``GL'' for brevity]
\ba
    \fbox{1}  & = \frac{1}{L}\int_0^{L/\lambda} \!\!\!\!\dd z~\Omega [M(\lambda,z)] \,,
    \\
    \fbox{2} & = - \frac{1}{\lambda} \Omega [M(\lambda, L/\lambda)]
    = - \frac{1}{\lambda}\Omega [M_0(L)]\,, 
    \\
    \fbox{3} & = \frac{1}{L/\lambda}
    \int_0^{L/\lambda}\!\!\!\!\dd z~\frac{\der}{\der \lambda} \Omega [M(\lambda, z)]
    \\
    & = \frac{1}{L/\lambda}
    \int_0^{L/\lambda}\!\!\!\! \dd z \bigg(
      \frac{\der \Omega }{\der M} + 
      \frac{\der \Omega }{\der M'} \der_z + 
      \frac{\der \Omega }{\der M''} \der_z^2
    \bigg) \frac{\der M}{\der \lambda} 
    \label{eq:B6q}
    \\
    & = \frac{1}{L}\int_0^{L/\lambda}\!\!\!\! \dd z \bigg[
    z \bigg( \frac{\der \Omega}{\der M}M' 
    + \frac{\der \Omega}{\der M'}M'' 
    + \frac{\der \Omega}{\der M''}M''' \bigg)
    \notag
    \\
    & \qquad 
    + \frac{\der \Omega}{\der M'} M' 
    + 2 \frac{\der \Omega}{\der M''} M''
    \bigg]
    \label{eq:B7q}
    \\
    & = \frac{1}{L}\int_0^{L/\lambda}\!\!\!\! \dd z \bigg[
    z \frac{\dd \Omega}{\dd z}
    + \frac{\der \Omega}{\der M'} M' 
    + 2 \frac{\der \Omega}{\der M''} M''
    \bigg] 
    \\
    & = \frac{1}{\lambda}\Omega[M_0(L)] 
    - \frac{1}{L}\int_0^{L/\lambda}\!\!\!\! \dd z~\Omega[M]  
    \notag
    \\
    & \qquad 
    + \frac{1}{L}\int_0^{L/\lambda}\!\!\!\! \dd z \bigg[
    \frac{\der \Omega}{\der M'} M' 
    + 2 \frac{\der \Omega}{\der M''} M''
    \bigg] \,.
\ea
In the step from \eqref{eq:B6q} to \eqref{eq:B7q}, we used the relation 
$\displaystyle \frac{\der M}{\der \lambda}=\frac{z}{\lambda}M'$. Now, 
plugging $\fbox{1}$, $\fbox{2}$ and $\fbox{3}$ 
into \eqref{eq:lambdazero} and taking the limit $\lambda\to 1$, we obtain
\ba
  \oint \bigg[
    \frac{\der \Omega}{\der M'} M' 
    + 2 \frac{\der \Omega}{\der M''} M''
    \bigg]\bigg|_{M=M_0} = 0 \,.
\ea
Recalling that $\Omega_{\rm GL}[M]$ is given by \eqref{eq:GL_NJL4}, this translates into
\ba
  \oint \big[
    2(\alpha_4+10\alpha_6M_0^2)(M_0')^2 + 4 \alpha_6 (M_0'')^2
  \big] = 0 \,,
\ea 
which reduces to $\displaystyle \oint g_1=0$ via integration by parts. 
This completes the proof. We stress that 
the stability of the condensate under dilatation has played 
an essential role here.

   \section{Variational analysis for the lowest eigenvalue spectrum of phonons}
   \label{sec:solut-schl-equat} 
   
   In this Appendix we apply a variational technique to the 
   eigenvalue problem \eqref{eq:schrodinger_eq} and show 
   that the energy of long-wavelength phonons is given, at leading order, 
   by \eqref{eq:dispersion_phonon}.  First of all, we note that, 
   among infinitely many energy levels that follow from \eqref{eq:dispersion_phonon}, 
   it is only the lowest level $E_0$ that matters for 
   the low-energy phonons. Then 
   it is easily seen that $E_0$ for given $k=(k_\bot,k_z)$ 
   is given by the formula
   \ba
   \begin{split}
     E_0(k) & = \min_{u\in \mathfrak{U}(k)} 
     \frac{1}{\oint  |u|^2}
     \oint \Big\{ f_1|\der_z u|^2 + f_2|\der_z^2 u|^2 
     \\
     & \quad 
     + (g_1+\der_z h_1)|\nabla_{\perp} u|^2 + g_2|\nabla_{\perp}^2 u|^2
     \\
     & \quad 
     + h_2 (\der_z^2\bar{u})(\nabla_{\perp}^2 u)
     + h_2 (\nabla_{\perp}^2 \bar{u})(\der_z^2u)
     \Big\} \,,
   \end{split}
   \label{eq:zeroE}
   \ea
   where $\mathfrak{U}(k)$ stands for the set of
   smooth functions of the form
   \ba
    \begin{split}
       u(\bm{x}) = \ee^{ik_z z}\ee^{i\bm{k}_{\perp} \cdot
       \bm{x}_{\perp}}\phi(z)\,, \quad 
       \\ 
       \phi(z+L)=\phi(z)\,, \quad 
       k_z,k_{\perp} \in\mathbb{R} \,.
    \end{split}
    \label{eq:ucond}
   \ea

  \subsection{\boldmath Eigenvalue with $k_\bot\ne 0$ and $k_z=0$}
  \label{sec:perp-direct-k_z=0}
  
  Now we consider $E_0$ for nonzero transverse momenta. 
  Substituting \eqref{eq:ucond} with $k_z=0$ into \eqref{eq:zeroE}, we get
  \ba
  \begin{split}
    &E_{0} (k_{\perp}) =  \min_{\phi} 
    \frac{1}{\oint |\phi|^2}
    \oint \Big[ 
    f_1 |{\phi}'|^2 + f_2 |{\phi}''|^2  
    \\
    &\qquad+ (k_{\perp}^2 \tilde g_1 + k_{\perp}^4  g_2) |\phi|^2   
    - k_{\perp}^2 h_2  (\bar{\phi''}
    \phi + \bar{\phi} \phi'' ) \Big] \,,
    \label{eq:Emin}
  \end{split}
  \ea
  where the primes denote derivatives by $z$ and 
  we defined $\tilde g_1\equiv g_1+h_1'$. 
  When $k_{\perp}=0$, the minimum is trivially $E_0=0$, which is achieved by
  an arbitrary constant solution: $\phi(z)=\phi_0\ne 0$. Now for
  sufficiently small $k_{\perp}$, we can expand  
  $\phi$ that corresponds to the minimum of \eqref{eq:Emin} 
  in perturbative series of $k_{\perp}$ as
  \ba 
    \phi(z) & = \phi_0 [1 + k_{\perp} \phi_1(z) + k_{\perp}^2 \phi_2(z) + \dots ] \,. 
  \ea
  Plugging this into \eqref{eq:Emin} yields 
  \ba
    E_{0}(k_{\perp}) &= 
    \min_{\phi}\ckakko{\beta_2[\phi_1] k_{\perp}^2 +
    \beta_3[\phi_i] k_{\perp}^3 +
    \beta_4[\phi_i] k_{\perp}^4 + \calO(k_{\perp}^5) }
    \notag 
    \\
    & \equiv \beta_{2*}k_\bot^2 + \beta_{3*}k_\bot^3 + \beta_{4*}k_\bot^4 
    + \calO(k_\bot^5) \,,
  \ea
  with 
  \begin{widetext}
  \ba
    \beta_2[\phi_1] &=\oint \ckakko{f_1
    |\phi'_1|^2 + f_2 |\phi''_1|^2 + \tilde{g}_1 } ,
    \\
    \beta_3[\phi_i] &= 
    \oint \ckakko{f_1 (\bar{{\phi}'_2} {\phi}'_1  +
    \bar{{\phi}'_1} {\phi}'_2 )   
    + f_2 ( \bar{{\phi}''_2} {\phi}''_1+
    \bar{{\phi}''_1} {\phi}''_2 ) + \tilde{g}_1 (\phi_1 + \bar{\phi}_1)
    - h_2 (\bar{\phi''_1} + \phi''_1 ) }  - 
    \beta_2[\phi] \oint (\phi_1 + \bar{\phi}_1) \,,
    \\
    \beta_4[\phi_i] &= \oint \Big\{ 
    f_1 ( |{\phi}'_2|^2 +
    \bar{{\phi}'_1} {\phi}'_3 + 
    \bar{{\phi}'_3} {\phi}'_1 ) + 
    f_2 ( |{\phi}''_2|^2 +
    \bar{ {\phi}''_1} {\phi}''_3 + \bar{{\phi}''_3} {\phi}''_1 ) 
    + \tilde{g}_1 ( |\phi_1|^2 + \phi_2 + \bar{\phi}_2 )  
    + g_2 
    \notag
    \\
    & \quad \quad \quad 
    - h_2 ( \bar{\phi''_1} \phi_1 +\bar{\phi}_1
    \phi''_1 +\bar{\phi''_2}+ \phi_2'' ) \Big\}  
    - \beta_2[\phi] \oint \mkakko{|\phi_1|^2  
    + \phi_2 +\bar{\phi}_2}
    - \beta_3[\phi]  \oint (\phi_1 + \bar{\phi}_1) \,.
  \ea
  \end{widetext}
  Since $E_0(k_\bot)$ is the minimal eigenvalue, the leading coefficient 
  $\beta_{2*}$ must be at the minimum as a functional of $\phi$, i.e.,  
  \ba
    \beta_{2*} = \min_{\phi_1} \oint \ckakko{f_1 |{\phi}'_1|^2 + f_2
    |{\phi}''_1|^2 + \tilde{g}_1 }\,. 
  \ea
  As $f_1$ and $f_2$ are positive functions, the minimum trivially occurs when 
  ${\phi}'_1=0$. Thus,  
  \ba
    \beta_{2*} = \oint \tilde{g}_1 = \oint g_1=0\,,
    \label{eq:beta2vanish}
  \ea 
  where the last equality is proved in Appendix \ref{sec:proof_g1=0}. 

  Next substituting the solution ${\phi}'_1=0$ into $\beta_3[\phi_i]$, 
  we readily find that the coefficient of $k_\perp^3$ vanishes:
  \ba
    \beta_{3*} = \min_{\phi_1 \in \mathbb{C}}
    \ckakko{ \oint \ckakko{ \tilde{g}_1  (\phi_1
    + \bar{\phi}_1)  } - \beta_{2*} \oint (\phi_1 + \bar{\phi}_1) }= 0\;.
  \ea
  
  Finally we come to the coefficient of $k_\bot^4$. Since it is the leading 
  nonzero term in $E_0(k_\bot)$, it must be minimized as a functional of $\phi$.  
  Substituting ${\phi}'_1=0$ into $\beta_4[\phi_i]$ and using 
  $\beta_{2*}=\beta_{3*}=0$, we obtain 
  \ba
   \begin{split}
    \beta_{4*} & = \oint g_2 + \min_{\phi_2} \oint
    \Big\{ f_1 | {\phi}'_2 |^2 + f_2 | {\phi}''_2 |^2 
    \\
    & \quad + \tilde{g}_1( \phi_2 + \bar{\phi}_2 ) - 
    h_2 (\bar{\phi''_2}+\phi_2'' )\Big\} \,. 
   \end{split}
  \ea
  Although it is not analytically tractable, one can in principle 
  determine $\phi_2$ that minimizes the integral by solving 
  the Euler-Lagrange equation 
  \ba
    - \mkakko{f_1 {\phi}'_2}' + (f_2{\phi}''_2)'' + \tilde{g}_1 - h''_2 =0 
    \label{eq:euloer-lagrange}
  \ea
  with periodic boundary conditions. 
  
  Summarizing above, the eigenvalue $E_0$ for transverse 
  momenta is given by
  \ba
    E_0(k_{\perp}) = \beta_{4*} k_{\perp}^4 + \calO(k_\bot^5)\,.  
  \ea

  \subsection{\boldmath Eigenvalue spectrum with $k_z \ne0$ and $k_{\perp}=0$}
  \label{sec:parall-direct-k_perp}
  Next we consider $E_0$ for parallel direction,
  \ba
  \begin{split}
   & E_0(k_z) \\
   &\quad= \min_{\phi} \frac{\oint \ckakko{ f_1 | ik_z \phi + \phi'|^2 
    + f_2| k_z^2\phi - 2i k_z \phi' - \phi''|^2 }}{\oint |\phi|^2} \,.
    \end{split}
    \label{eq:E0ddd}
  \ea
  When $k_z=0$, the minimum is trivially $E_0=0$, corresponding 
  to a constant solution $\phi(z)=\phi_0\ne
  0$. Now, for sufficiently small $k_z$, one can expand $\phi$ that
  achieves the minimum of \eqref{eq:E0ddd} in a perturbative series 
  of $k_z$ as
  \ba
   \phi(z) & = \phi_0  [ 1 + k_z \chi_1(z) + k_z^2 \chi_2(z) +
   \dots ] \,.   
  \ea

  Substituting this expansion, we find
  \ba
    E_0(k_z) &=  \min_{\phi} \ckakko{\gamma_{2}[\chi_1] k_z^2 +
    \calO(k_z^3) }\,,
    \intertext{with}
    \gamma_2[\phi_1]& \equiv \oint \ckakko{f_1 |1 - i \chi_1'|^2 + f_2
    |\chi_1''|^2} \,.
  \ea
  Since $E_0(k_z)$ is the minimal eigenvalue, $\chi_1$ must be chosen 
  so as to minimize $\gamma_2[\chi_1]$. Thus, $\chi_1$
  satisfies the Euler-Lagrange equation
  \ba
   \frac{\delta \gamma_2[\chi_1]}{\delta \bar{\chi}_1} 
   = - i f_1' - (f_1 \chi_1')'
    +(f_2 \chi_1'')'' = 0 \,, \label{eq:phi1}
  \ea
  with periodic boundary conditions. 
  Using the solution $\chi_{1*}$, the eigenvalue at small $k_z$ 
  is finally obtained as 
  \ba
    E_{0}(k_z) & = \gamma_{2*} k_z^2 + \calO(k_z^3) \,,
  \ea
  with $\gamma_{2*} \equiv \gamma_{2}[\chi_{1*}]$.

\subsection{\boldmath Absence of $k_zk_\perp^2$ term in $E_0$}
\label{app:nok3term}

 We have shown the leading behavior of $E_0$ for \mbox{$k_\perp\neq0$}, \mbox{$k_z=0$}, and for $k_\perp\neq0, k_z=0$ in the previous subsections.  
 Here we show that $E_0$ does not have a $k_zk_\perp^2$ term, when both $k_z$ and $k_\perp$ are nonzero.
 
 Since $H_u$ in \eqref{eq:Hu} is invariant under $z\to -z$, 
 the eigenvalues with the momentum $k_z$ and $-k_z$ degenerate,
   which implies the absence of the $k_zk_\bot^2$ term.
   To see this explicitly, let us decompose $H_u$ into an unperturbed 
   part $H_0$ and a perturbation $V$:
   \ba
   H_0 &\equiv - \der_z (f_1 \der_z)+\der_z^2 (f_2\der_z^2) \,,
   \\
   V &\equiv 
       - g_1 \nabla_{\perp}^2 + g_2 \nabla_{\perp}^4   - (\der_z h_1)  
       \nabla_{\perp}^2 + \nabla_{\perp}^2 \ckakko{h_2,\der_z^2}_+ \,.
   \ea
The lowest eigenvalue state for the unperturbed part 
can be expanded as 
\ba
       u(\bm{x}) = \ee^{ik_z z}\ee^{i\bm{k}_{\perp} \cdot
       \bm{x}_{\perp}}\!\phi_0[1+k_z\chi_1+k_z^2\chi_2+\cdots]\,, 
\ea
where $\chi_1$ satisfies \eqref{eq:phi1}. Since $f_1$ and $f_2$ are 
invariant under $z\to L-z$, the solution of \eqref{eq:phi1}, $\chi_1(z)$, 
is an odd function, $\chi_1(L-z)=-\chi_1(z)$. Due to the periodic 
boundary condition, $\chi_1(0)=\chi_1(L)=0$.

The eigenvalue for the unperturbed part is obtained as
\ba
  E_0(k_z)=  \frac{1}{\oint |u|^2}
  \oint \bar u H_0 u =  \gamma_{2*}k_z^2 +\mathcal{O}(k_z^3)\,.
\ea
Since $V$ is of order $\nabla_\perp^2$, i.e., $k_\perp^2$ in momentum space, the first-order correction $\delta E_0$ gives 
the contribution of order $k_\perp^2$, which is given by the expectation value of $V$ for $u$:
\ba
  \begin{split}
    \delta E_0 &=  \frac{1}{\oint |u|^2}\oint \bar u V u 
    \\
    &= k_zk_\perp^2\oint  \Big\{\tilde{g}_1(\chi_1+\bar\chi_1)  -   
    h_2(\chi_1''+\bar\chi_1'')\Big\}
    +  \mathcal{O}(k_z^2k_\perp^2) \,.
    \label{eq:deltaE0}
  \end{split}
\ea
In the second line, we used \eqref{eq:beta2vanish} and integration by parts.
The integrand is  odd under $z\to L-z$ because $\chi_1(L-z)=-\chi_1(z)$, $\tilde{g}_1(L-z)=\tilde{g}_1(z)$ and $h_2(L-z)=h_2(z)$. 
Therefore, the integral in the second line of \eqref{eq:deltaE0} vanishes, which proves the absence of the $k_zk_\perp^2$ term in $E_0$.

\bibliography{paper_v2.bbl}
\end{document}